\documentclass[a4paper,11pt,feynmf]{article}
\usepackage{graphicx}
\usepackage{amsmath,amssymb}
\usepackage{graphics,color,array,calc,rotating,epsfig,psfrag}
\numberwithin{equation}{section}
\usepackage{cite}
\usepackage{bm}
\usepackage{dcolumn}
\usepackage{float}
\usepackage{subfigure}
\usepackage{amsfonts}
\usepackage[mathscr]{eucal}
\def\be{\begin{equation}} \def\ee{\end{equation}}
\def\bea{\begin{eqnarray}} \def\eea{\end{eqnarray}}

\begin{document}
	
	%
	\baselineskip 18pt%
	\begin{titlepage}
		\vspace*{1mm}%
		\hfill%
		\vspace*{15mm}%
		\hfill
		\vbox{
			\halign{#\hfil         \cr
			} 
		}  
		\vspace*{20mm}
		
		\begin{center}
			{\large {\bf
					Transverse and 	non-boost longitudinal   expansion of    (2+1)dimensional relativistic   ideal-hydrodynamics flow 	
					 in heavy ion collisions.								 				
			}}\\
			\vspace*{5mm}
			{  M.~Karimabadi, A.~F.~Kord\footnote{a.f.kord@hsu.ac.ir}, B.~Azadegan }\\
			\vspace*{0.2cm}
			{$^{}$ Department of Physics, Hakim Sabzevari University, P.O. Box 397, Sabzevar, Iran}\\
			\vspace*{1cm}
		\end{center}
		\begin{abstract}
		This study investigates the evolution of quark gluon plasma (QGP) within a generalized Bjorken flow framework. The medium under consideration is assumed to possess a finite transverse size and to expand both radially and along the beam axis. However, we assume that the boost invariance of longitudinal expansion is broken.
		
		To be more specific, we generalize the Bjorken solution to include the acceleration and transverse expansion of the fluid. We analytically study the (2 + 1) dimensional longitudinal acceleration expansion of hot and dense quark matter, applying a perturbation approach to solve the relativistic hydrodynamics equations. This procedure enables us to obtain exact algebraic expressions for fluid velocities and energy densities in both transverse and longitudinal directions.
			
		To simplify our calculations, we assume that the fluid is produced in central collisions, and therefore, we consider azimuthal symmetry. We compare the radial velocity and correction energy density with those obtained from the Gubser model. 
				
		Furthermore, we determine the fluid's acceleration parameter and longitudinal correction energy density, which exhibits a Gaussian distribution.

		\end{abstract}
		
	\end{titlepage}
	
\section{Introduction}
Based on experimental data obtained from relativistic Heavy Ion Collisions (HICs) conducted at RHIC and LHC, a distinct form of hot and dense nuclear matter is generated during the initial stages of collisions, commonly referred to as Quark-Gluon Plasma (QGP). It has been observed that QGP exhibits characteristics of a strongly-coupled, nearly perfect fluid. The application of relativistic hydrodynamics to describe the QGP phase has yielded promising results in heavy-ion collision experiments \cite{a1,a2,a3,a4,a5,a6}.

The Bjorken flow model is a straightforward scenario that characterizes the typical motion of partons following a collision \cite{a1new}. This model is founded on certain assumptions, including boost invariance along the beam line, as well as translation and rotation invariance in the transverse plane. Consequently, all relevant quantities can be expressed as functions of the proper time $\tau$ in the $(\tau,x_\perp,\phi,\eta)$ Milne coordinate system. By considering the aforementioned symmetries, along with the system's invariance under reflection, one can determine the four-velocity profile, which is given by $u^\mu=(1,0,0,0)$ in the Milne coordinate system.

 The Bjorken model, even in central collisions, is subject to two issues. Firstly, the model predicts that the radial flow $(u_{x_\perp})$ is zero due to translation invariance in the transverse plane. However, this symmetry is not realistic as the size of colliding nuclei is limited, which may result in misleading subsequent hydrodynamical flow, on  which much of heavy-ions phenomenology depends.  Secondly, the model predicts a flat rapidity distribution of final particles, which is inconsistent with observations at RHIC, except for a limited region around mid-rapidity. In realistic collisions, boost and translation invariance are violated, and a model that is more faithful and not far from the accelerationless Bjorken picture should be investigated. Several attempts have been made to generalize the Bjorken model, such as those presented in \cite{a7}. Some of these attempts have included accelerating solutions of relativistic fluid dynamics to obtain more realistic estimations, as seen in \cite{a8}-\cite{a11}. A recent paper based on accelerating hydrodynamic description can be found in \cite{a12,aa12}.

  The objective of our research is to extend the Bjorken model in a generalized manner.  Our approach assumes the breaking of translation and boost invariance, while the rotational symmetry around the beam line is maintained. Our investigation focuses on central collisions, wherein we present an analytical solution for the transverse and longitudinal expansions of a plasma, utilizing perturbation theory. Our model is closely related to the Bjorken model, and we assume that the medium is formed rapidly following the collisions, with generalized Bjorken transverse and longitudinal expansions. Our aim is to derive solutions that represent the resistive relativistic hydrodynamic extension of the one-dimensional generalized Bjorken flow, along both the $z$ and $x_\perp$ directions.

This study focuses on the specific scenario of a (2 + 1) dimensional fluid that undergoes non-boost-invariant expansion, while also experiencing radial expansion in the transverse plane. Through analytical means, a novel solution of relativistic hydrodynamics in (2+1) dimensions is derived, which is dependent upon three variables: proper time $(\tau)$, transverse coordinate $(x_\perp)$, and rapidity $(\eta)$.

   The present paper is structured as follows. Section 2 provides an exposition of the ideal relativistic hydrodynamic framework, specifically in the context of a plasma. Subsequently, we present our perturbative approach and derive analytical solutions. The findings are then discussed in section 3, where we provide a comprehensive analysis of the general results. Finally, the last section serves to summarize the conclusions drawn from our study and highlight potential avenues for future research.

 \section{Ideal relativistic  fluid expansion}
 In this section, we provide a succinct overview of our formalism for characterizing the assessment of QGP matter. Additionally, we take into account the rotational symmetry of the medium with respect to the beam line, which is applicable to central collisions. Consequently, we posit that all relevant quantities are dependent solely on the transverse radial coordinate $x_\perp$, the proper time $\tau$, and the rapidity $\eta$ in the Milne $(\tau, r, \phi, \eta)$ coordinate system.

In this paper, we examine the scenario of an ideal non-resistive plasma consisting of massless particles. Additionally, we propose the inclusion of a thermodynamic equation of state (EOS) that assumes the pressure to be directly proportional to the energy density, expressed as $P = 1/3\epsilon$, in order to close the set of equations. We proceed to present the energy-momentum conservation equations for an ideal fluid.

The energy-momentum conservation equations for an ideal fluid can be expressed in a covariant form, which is given by:
\begin{eqnarray}\label{1}
d_\mu T^{\mu\nu}=0,
\end{eqnarray}
where
\begin{eqnarray}\label{2}
T^{\mu\nu}&=&(\epsilon+P)u^\mu u^\nu+Pg^{\mu\nu}.
\end{eqnarray}

The energy density and pressure of the fluid are denoted by $\epsilon$ and $P$, respectively. In a flat spacetime, the metric tensor is represented by $g_{\mu \nu}=diag \lbrace-,+,+,+\rbrace$. Additionally, the four velocity of the single fluid, denoted as $u^{\mu}$ with the constraint $u_{\mu}u^{\mu}=-1$, can be expressed as $u^\mu=\gamma(1, \vec v)$, where $\gamma=\frac{1}{\sqrt{1-v^2}}$.

The covariant derivative is expressed in Eq.(\ref{1}) as follows:
\begin{eqnarray}\label{2b}
d_p A^{\mu\nu}&=&\partial_p A^{\mu\nu}+\Gamma^\mu_{p m} A^{m \nu}+\Gamma^\nu_{p m} A^{ \mu m}.
\end{eqnarray}\label{2bb}
The symbols $\Gamma^i_{j k}$ denote the Christoffel symbols. They are:
\begin{eqnarray}\label{3b}
\Gamma^i_{jk}=\frac{1}{2}g^{im}\left(\frac{\partial g_{mj}}{\partial x^k}+\frac{\partial g_{mk}}
{\partial x^j}-\frac{\partial g_{jk}}{\partial x^m}\right).
\end{eqnarray}

One can express the conservation equations by projecting $d_\mu T^{\mu\nu}=0$ along the longitudinal and transverse directions relative to $u^\mu$.

\begin{eqnarray}\label{5}
	u_\nu( d_\mu T_{matter}^{\mu\nu}=0 )
	&\rightarrow& D\epsilon+(\epsilon+P)\Theta=0,
	\\\label{6}
	\Delta_{\alpha \nu} ( d_\mu T_{matter}^{\mu\nu}=0 )
	&\rightarrow&(\epsilon+P)Du_\alpha+\nabla_\alpha P=0,
\end{eqnarray}
where 
\begin{eqnarray}
	D=u^\mu d_\mu,\ \Theta=d_\mu u^\mu,\ \nabla^\mu=d^\mu+u^\mu D,\ \Delta^\alpha_\nu=g^\alpha_\nu+u^\alpha u_\nu.
\end{eqnarray}

\subsection{Method}
In the realm of central collisions involving two nuclei, it is hypothesized that the matter undergoing expansion exhibits azimuthal symmetry. This postulation prompts the examination of the four-vector velocity of the  matter, as follows:
 \begin{eqnarray}
\label{3}
&& u^
\mu=\gamma(1,  v_{x_\perp}, 0,v_z )=(\cosh K \cosh Y,\sinh K,0,\cosh K\sinh Y). 
\end{eqnarray}
The present study considers the transverse and longitudinal fluid rapidities, denoted by $Y$ and $K$, respectively, with $v_z = \tanh Y$ and $v_{x_\perp}=\frac{\tanh K}{\cosh Y}$. To facilitate our analysis, we make the assumption that $K$ is solely dependent on $(x_\perp,t)$, while $Y$ is dependent on $(z,t)$.

The utilization of Milne coordinates is deemed more expedient in comparison to the conventional Cartesian coordinates.
\begin{eqnarray}
	(\tau, x_\perp, \phi, \eta)&=&\left(\sqrt{t^2-z^2},x_{\perp},\phi,\frac{1}{2}ln\frac{t+z}{t-z}\right).
\end{eqnarray}
 The metric is given by:
\begin{eqnarray}
	g^{\mu\nu}=diag(-1, 1, 1/x^2_\perp, 1/\tau^2),  \  \  \  \ g_{\mu\nu}=diag(-1, 1, x^2_\perp, \tau^2).
\end{eqnarray}
By utilizing Milne coordinates, the Christoffel symbols can be readily derived. The sole non-zero symbols are as follows: $\Gamma^\tau_{\eta\eta}=\tau$, $\Gamma^\eta_{\tau\eta}= 1/\tau$, $\Gamma^{x_\perp}_{\phi\phi} =-x_\perp$, and $\Gamma^\phi_{x_\perp\phi}=1/x_\perp$. 
Furthermore, in the Milne coordinate system, the four-vector velocity is expressed as follows:
\begin{eqnarray}
\label{4}
&& u^
\mu=(\cosh K\cosh(Y-\eta),\sinh K,0,\cosh K\sinh(Y-\eta)/\tau)=\bar{\gamma}(1,v_\perp,0,0,v_\eta/\tau), \ \
\end{eqnarray}
 where $\bar{\gamma}=\cosh K(\tau,x_\perp) \cosh (Y(\tau,\eta)-\eta), \ v_{x_\perp} =\frac{\tanh K}{ \cosh (Y-\eta) }$  and $v_\eta=\tanh (Y-\eta)$.

The aforementioned assumptions facilitate the rewriting of the conservation equations in Milne coordinate. The energy and Euler equations  can be expressed as follows:

\begin{eqnarray}
\label{7}
 &&(\epsilon+P)(v_\eta\tau\partial_{\tau}Y + \partial_\eta Y)+(\tau\partial_\tau \epsilon+v_\eta\partial_\eta \epsilon+ \tau v_{x_\perp} \partial_{x_\perp}\epsilon)\nonumber\\&&  +(\epsilon+P)(\tau\tanh K\partial_\tau K+\frac{\tau}{\cosh(Y-\eta) }(\partial_{x_\perp}K)+\frac{\tau}{x_\perp} v_{x_\perp})=0, \\&&
 \label{8}
 (\epsilon+P)(\cosh K \partial_{\tau}K+\frac{\sinh K}{\cosh (Y-\eta)} \partial_{x_\perp}K)+\nonumber\\&& 
 (\frac{v_\eta \sinh K}{\tau}\partial_\eta p+\sinh K\partial_{\tau}P+\frac{\cosh K}{\cosh(Y-\eta)}\partial_{x_\perp}P)=0,  \\&&
 \label{9}
 (\epsilon+P)(\tau\partial_\tau Y+v_\eta \partial_\eta Y+\tau v_\eta \tanh k \partial_\tau K +\tau v_\eta v_\perp \tanh K \partial_{x_\perp} K )\nonumber\\&&  +(v_\eta \tau \partial_\tau P+v_\eta^2\partial_\eta P+\frac{1}{\cosh K^2\cosh^2(Y-\eta)}\partial_\eta P
 )+
 \tau v_\eta v_{x_\perp}\partial _{x_\perp}P=0.
 \end{eqnarray}
 
 In the present study, we shall examine perturbative solutions of the aforementioned conservation equations. The desired quantities will be expressed as power series in $\lambda_1$. The convergence of the series is not a primary concern, as it has been demonstrated that that this method accurately describes the physical system under investigation. Within the framework of the perturbative approach, it is postulated that specific assumptions are made in the following manner:

\begin{eqnarray}
\label{10}
&&\epsilon(\tau,x_{\perp},\eta)=\epsilon^0(\tau)+ \lambda_1 \epsilon^{(1)}(\tau,x_{\perp},\eta)+\lambda_1^2\epsilon^{(2)}(\tau,x_{\perp},\eta)+...\\&& \label{11}
Y(\tau,\eta)=\eta +\lambda_1 Y^{(1)}(\tau,\eta)+ \lambda_1 ^2 Y^{(2)}(\tau,\eta)+...\\&& \label{12}
K(\tau,x_\perp)=\lambda_1 k^{(1)}(\tau,x_\perp)+\lambda_1^2 K^{(2)}(\tau, x_\perp)+....
\end{eqnarray}
 In this context, the symbol $\lambda_1$ represents an expansion parameter that will ultimately be assigned a value of one upon completion of the calculations.
 To address the conservation equations under the aforementioned assumptions, it is necessary to commence by simplifying certain terms in Eqs~(\ref{7}-\ref{9}):

\begin{eqnarray}
\label{13}
&&\cosh(Y-\eta)\simeq  1+1/2\lambda_1 ^2 Y^{(1)}+... ,\  1/\cosh(Y-\eta )\simeq 1+...,\  v_\eta\simeq \lambda_1 Y^{(1)} +\lambda_1 ^2 Y^{(2)}+...\nonumber\\ && \sinh K \simeq \lambda_1 K^{(1)} +\lambda_1 ^2 K^{(2)}+..., \ \cosh K\simeq 1+1/2 \lambda_1 ^2 K^{(1)}+..., \ v_\perp\simeq \lambda_1 K^{(1)} +\lambda_1^2 K^{(2)} +.... \nonumber\\ &&
\end{eqnarray}

 We have retained the terms up to second order of $\lambda_1$ in our analysis. Additionally, we have taken into account that the fluid is highly relativistic, thereby rendering the rest mass contributions to the equation of state (EOS) negligible. Consequently, the pressure can be expressed as a simple proportionality to the energy density, i.e., $P = c_s^2\epsilon = 1/3\epsilon$, where $c_s = 1/3$ denotes the speed of sound. Upon substituting the aforementioned assumptions into the conservation equations (\ref{7}-\ref{9}) and identifying the powers of $\lambda_1$, a series of equations can be obtained.
 
Thus, it is possible to express the equation to the first degree of $\lambda_1$ as follows:
\begin{eqnarray}
\label{14}
&&\tau\partial_{\tau}\epsilon^{(0)} +\frac{4}{3}\epsilon^{(0)} =0, 
\nonumber\\ && Y^{(0)}=\eta, \ \ K^{(0)}=0.
\end{eqnarray} 
The standard result for Bjorken flow, namely $\epsilon^0=\epsilon_c(\frac{\tau_0}{\tau})^{4/3}$, has been obtained herein.

The equations resulting from the identification of terms proportional to first order in $\lambda_1$ are presented as follows:
\begin{eqnarray}
\label{15}
&&\epsilon^0 Y^{(1)} +  \epsilon^{(1)}+\frac{3}{4}\tau\partial_{\tau} \epsilon^{(1)}+\tau \epsilon^0 \partial_{x_\perp} K^{(1)}+\frac{\tau}{x_\perp}\epsilon^0 K^{(1)}=0, \\ &&
\label{16}
4\epsilon^0 \partial_{\tau} K^{(1)}+K^{(1)}\partial_{\tau}\epsilon^0+ \partial_{x_\perp}\epsilon^{(1)}=0,  \\ &&
\label{17}
4\epsilon^0 \partial_{\tau}( \tau Y^{(1)})+Y^{(1)}\tau\partial_{\tau}\epsilon^0+ \partial_{\eta}\epsilon^{(1)}=0.
\end{eqnarray}

Furthermore, the equations resulting from the identification of terms that are proportional to the second order in $\lambda_2$ are expressed as follows:
\begin{equation}
\label{15a}
	\begin{aligned}
		& \epsilon^2 + \frac{K^{2}\; \epsilon^0\; \tau}{x_{\perp}}+ \frac{K^{1}\; \epsilon^1\; \tau}{x_{\perp}}+ \epsilon^1\;\tau\; \partial_{x_{\perp}} K^1+ \epsilon^0\;\tau\; \partial_{x_{\perp}} K^2\;+ \epsilon_1\; \partial_\eta Y^1+ \epsilon_0\; \partial_\eta Y^2 \\
		&+ K^1 \epsilon^0 \tau \partial_\tau K^1\;+\;+ Y^1 \epsilon^0 \tau \partial_\tau Y^1\;+\; \frac{3}{4} Y^1\; \partial_\eta \epsilon^1\;+\; \frac{3}{4} K^1\;\tau \; \partial_{x_{\perp}} \epsilon^1\;+\; \frac{3}{4} \;\tau \; \partial_\tau \epsilon^2=0,
	\end{aligned}
\end{equation}

\begin{equation}
K^2 \partial_\tau \epsilon^0\;+\; 4 K^1 \epsilon^0 \partial_{x_{\perp}} K^1\;+\; 4 \epsilon^1 \partial_\tau K^1\;+\; 4 \epsilon^0 \partial_\tau K^2\;+\; \partial_{x_{\perp}} \epsilon_2\;+\;K^1 \partial_\tau \epsilon_1=0,
\end{equation}

\begin{equation}
\begin{aligned}
	&4 \epsilon^1 Y^1\;+\;4 \epsilon_0 Y^2\;+\;\tau Y_2 \partial_\tau \epsilon^0\;+\;4 \epsilon^0 Y_1 \partial_\eta Y^1\;+\;4\epsilon_1 \tau \partial_\tau Y^1\;+\;4\epsilon^0 \tau \partial_\tau Y^2\\
	&+\partial_\eta \epsilon^2\;+\;\tau Y^1 \partial_\tau \epsilon^1=0\;.
\end{aligned}
\end{equation}

\subsection{First order expansion equations}

In the subsequent section, the conservation equations will be solved up to the first order expansion, as presented in Eqs~(\ref{15}-\ref{17}). The method of separation of variables will be employed to obtain the ordinary differential equations. To achieve this, we make the assumption:
\begin{eqnarray}
	\label{18}
	&&\epsilon^{(1)}(\tau,x_\perp,\eta)=\epsilon^1_{x_\perp}(\tau,x_\perp)+\epsilon^1_{\eta}(\tau,\eta).
\end{eqnarray}
The aforementioned formula can be utilized to transform Eqs~(\ref{15}-\ref{17}) into simplified expressions, which are presented as follows:
\begin{eqnarray}
	\label{19}
	&&\epsilon^{(1)}_{x_\perp}+\frac{3}{4}\tau\partial_{\tau} \epsilon^{(1)}_{x_\perp}+\tau \epsilon^0 \partial_{x_\perp} K^{(1)}+\frac{\tau}{x_\perp}\epsilon^0 K^{(1)}=n^2,
	\\
	&&
	\label{20}
	4\epsilon^0 \partial_{\tau} K^{(1)}+K^{(1)}\partial_{\tau}\epsilon^0+ \partial_{x_\perp}\epsilon^{(1)}_{x_\perp}=0, \\ &&
	\label{21}
	\epsilon^0\partial_{\eta} Y^{(1)} +  \epsilon^{(1)}_\eta+\frac{3}{4}\tau\partial_{\tau} \epsilon^{(1)}_\eta=-n^2, \\ &&
	\label{22}
	4\epsilon^0 \partial_{\tau}( \tau Y^{(1)})+Y^{(1)}\tau\partial_{\tau}\epsilon^0+ \partial_{\eta}\epsilon^{(1)}_\eta=0.
\end{eqnarray}
The equations denoted by (\ref{19}) and (\ref{20}) are dependent solely upon the proper time $\tau$ and the transverse coordinate $x_\perp$, and serve to describe the dynamic evolution of the fluid in the transverse direction. Additionally, the equations represented by (\ref{21}) and (\ref{22}) are dependent solely upon the proper time $\tau$ and the space rapidity $\eta$, and serve to describe the dynamic evolution of the fluid in the longitudinal direction. It should be noted that the real number denoted by $n$ must be identified based on physical conditions.

  The amalgamation of Equations (\ref{19}) and (\ref{20}) results in a partial differential equation that solely comprises the variable of $K^{(1)}$, as given by. 
\begin{equation}
\label{23}
\partial_{x_{\perp}}^{2} K^{(1)}-3\partial_\tau^{2}K^{(1)}+\frac{\partial_{x_\perp} K^{(1)}}{x_\perp}+\frac{\partial_{\tau} K^{(1)}}{\tau}- K^{(1)}(\frac{1}{x_\perp ^2}+\frac{1}{\tau^2})=0,
\end{equation}
The aforementioned partial differential equation can be solved through the method of separation of variables. The resulting general solution is as follows:
\begin{equation}
\label{24}
K^{(1)}(x_\perp,\tau)=\sum _k (c_1^k J_1(k x_\perp)+c_2^k Y_1(k  x_\perp)) \left(\tau ^{2/3} \left(c_3^k J_{\frac{1}{3}}\left(\frac{k \tau }{\sqrt{3}}\right)+c_4^k Y_{\frac{1}{3}}\left(\frac{k \tau }{\sqrt{3}}\right)\right)\right).
\end{equation}
The Bessel functions $J_1, Y_1, J_{\frac{1}{3}}$, and $Y_{\frac{1}{3}}$ are of interest in this study. For each value of $k$, there exist four integration constants, which are typically determined by initial conditions. Alternatively, the integration constants may be reduced by imposing the initial conditions $K^{(1)}(x_\perp=0,\tau)=0$ and $K^{(1)}(x_\perp,\tau \rightarrow \infty )=0$. It is important to note that, up to the first order expansion of the energy and Euler equations, $u_\perp=sinhK\simeq K^{(1)}$. Consequently, $c_2^k=c_4^k=0$, yielding:
 \begin{equation}
 \label{25}
 K^{(1)}(x_\perp,\tau)=\sum _k c^k J_1(k x_\perp) \tau
 ^{2/3}  J_{\frac{1}{3}}(\frac{k \tau }{\sqrt{3}}).
 \end{equation}

To determine the integration constants $c_k$, it is necessary to have knowledge of the space-time profile of the radial velocity $u_\perp(\tau,x_\perp)$ at $\tau=\tau_0$. To achieve this, we will utilize the analytic conformal four velocity $u_\mu(\tau,x_\perp,\phi,\tau)$ discovered by Gubser \cite{a35}. The Gubser fluid velocity ($u^\mu$) has only two non-zero components, namely $u^\tau$ and $u^\perp$, which describe the boost-invariant longitudinal expansion and the transverse expansion, respectively. These components are expressed as follows:
  \begin{equation}
 \label{26a}
 u^{\perp}(x_\perp,\tau)=\frac{q x_\perp}{\sqrt{1+g^2(x_\perp,\tau)}}.
 \end{equation}
 
 \begin{equation}
 \label{26b}
 u^{\tau}(x_\perp,\tau)=\frac{1+q^2 x^2_\perp+q^2 \tau^2}{2q\tau\sqrt{1+g^2(x_\perp,\tau)}}.
 \end{equation}
 The function $g(x_\perp,\tau)$ is hereby defined as follows:
 \begin{equation}
 \label{27}
 g(x_\perp,\tau)=\frac{1+q^2 x_\perp^2-q^2 \tau^2}{2 q \tau}.
 \end{equation} 
 
  Subsequently, the conformal hydrodynamic solution shall be employed as the initial condition at $\tau_0$. Specifically, it is postulated that the fluid adheres to the characteristics of a Gubser fluid during the initial proper time $\tau_0$. Consequently, the radial fluid velocity field at $\tau_0$ can be represented by a profile.  
 \begin{equation}
\label{26c}
u^{(1)}(x_\perp,\tau_0)=\frac{q x_\perp}{\sqrt{1+g^2(x_\perp,\tau_0)}}.
\end{equation}

It is postulated that our proposed solution, as denoted by Eq.(\ref{25}), is equivalent to Gubser's radial velocity solution, represented by Eq.(\ref{26c}), when evaluated at $\tau=\tau_0$.
\begin{equation}
\label{255}
u^{(1)}(x_\perp,\tau_0)= \frac{q x_\perp}{\sqrt{1+g^2(x_\perp,\tau_0)}}= \sum _k c^k J_1(k x_\perp) \tau_0^{2/3}  J_{\frac{1}{3}}(\frac{k \tau_0 }{\sqrt{3}}).
\end{equation}
The determination of the coefficients $c^k$ can be achieved through the utilization of the orthogonality of Bessel functions. These coefficients are expressed as follows:
\begin{equation}
\label{28}
c^k=\frac{2}{a^2(J_2(\beta_{1k}))^2 J_{\frac{1}{3}}(\beta_{1k}\frac{\tau_0}{a \sqrt{3}})}\int_0^a \frac{q x_\perp^2}{\sqrt{1+g^2(x_\perp,\tau_0)}} J_1(\beta_{1k}\frac{x_\perp}{a}) dx.
\end{equation}
The $k$th zero of $J_1$ is denoted by $\beta_{1k}$. It should be noted that in the aforementioned equation, $k$ is equivalent to the ratio of $\beta_{1k/a}$ ($k = \beta_{1k}/a$).
Ultimately, the transverse fluid velocity is expressed as:
 \begin{equation}
 \label{25a}
u^\perp =u^{(1)}(x_\perp,\tau)= \sum _k\Bigg[\frac{2}{a^2(J_2(\beta_{1k}))^2 J_{\frac{1}{3}}(\beta_{1k}\frac{\tau_0}{a \sqrt{3}})}\int_0^a \frac{q x_\perp^2}{\sqrt{1+g^2(x_\perp,\tau_0)}} J_1(\beta_{1k}\frac{x_\perp}{a}) dx_\perp \Bigg] J_1(k x_\perp) \tau^{2/3}  J_{\frac{1}{3}}(\frac{k \tau }{\sqrt{3}}).
 \end{equation}

Based on our assumptions, the total energy density up to the first order in $\lambda_1$ can be given by:
\begin{equation}
\label{28a}
\epsilon(\tau,x_\perp,\eta)=\epsilon^0 +\epsilon^1_{x_\perp}+\epsilon^1_\eta. 
\end{equation}
The density energy distribution in the transverse plane is denoted by $\epsilon^1_{x_\perp}$, while the density energy distribution in the longitudinal direction is denoted by $\epsilon^1_{\eta}$. To determine the transverse energy density distribution $\epsilon^1_{x_\perp}$, we combine equations (\ref{19}) and (\ref{20}). This yields the following partial differential equation:
\begin{equation}
\label{29}
3\partial_\tau^2 \epsilon_{x_\perp}^{(1)}-\partial_{x_\perp}^2 \epsilon_{x_\perp}^{(1)}+\frac{7 \partial_\tau \epsilon_{x_\perp}^{(1)}}{\tau}-\frac{\partial_{x_\perp}\epsilon_{x_\perp}^{(1)}}{x_\perp}=0,
\end{equation}
The equation (\ref{29}) may be solved through the method of separation of variables. The  general solution that is as follows:
\begin{equation}
\label{30}
\epsilon_{x_\perp}^{(1)}(x_\perp,\tau)=\sum _k\left (c_1^{\prime k} J_0(k  x_\perp)+c_2^{\prime k} Y_0(k  x_\perp)\right) \left(\tau ^{-2/3} \left(c_3^{\prime k} J_{\frac{2}{3}}\left(\frac{k \tau }{\sqrt{3}}\right)+c_4^{\prime k} Y_{\frac{2}{3}}\left(\frac{k \tau }{\sqrt{3}}\right)\right)\right).
\end{equation}
To ensure consistency between the above solution and the transverse fluid velocity equation (\ref{25a}), it is necessary to set $c_{2,4}^{\prime k}=0$. Consequently, the solution can be expressed as follows:
\begin{equation}
\label{31a}
\epsilon_{x_\perp}^{(1)}(x_\perp,\tau)=\sum _k c^{\prime k} J_0(k x_\perp) \tau
^{-2/3} J_{\frac{2}{3}}(\frac{k \tau }{\sqrt{3}}).
\end{equation}

Furthermore, it is posited that the fluid being studied exhibits characteristics akin to those of Gubser's inviscid hydrodynamic fluid at the specific time of $\tau=\tau_0$. As a result, the value of $\epsilon_{x_\perp}^{(1)}(x_\perp,\tau_0)$ can be expressed as:
\begin{equation}
\label{32}
\epsilon_{x_\perp}^{(1)}(x_\perp,\tau_0)=\epsilon_g(x_\perp,\tau_0)-\epsilon_0(\tau_0),
\end{equation}
where $\epsilon_g(x_\perp,\tau)$ is given by~\cite{a35}:
\begin{equation}
\label{33}
\epsilon_g(x_\perp,\tau)=\frac{\hat{\epsilon}_0}{\tau^{4 / 3}} \frac{(2 q)^{8 / 3}}{\left[1+2 q^2\left(\tau^2+x_{\perp}^2\right)+q^4\left(\tau^2-x_{\perp}^2\right)^2\right]^{4 / 3}}.
\end{equation}
Gubser's energy density for a conformal inviscid fluid is denoted by $\epsilon_g$, while $\hat{\epsilon}$ and $q$ are two constants. The reciprocal of $q$ is directly proportional to the transverse size of the plasma. By utilizing Eqs.~(\ref{33}, \ref{32}, \ref{31a}), and the orthogonality of Bessel functions, one can derive the coefficients in (\ref{31a}). These coefficients are:
\begin{equation}
\label{34}
c^{\prime k}=\frac{2}{a^2(J_1(\beta_{0k}))^2 J_{\frac{2}{3}}(\beta_{0k}\frac{\tau_0}{a \sqrt{3}})}\int_0^a x_\perp (\epsilon_g(x_\perp,\tau_0)-\epsilon_0(\tau_0)) J_0(\beta_{0k}\frac{x_\perp}{a}). dx_\perp
\end{equation}
$\beta_{0k}$ being the $k$th zero of $J_0$; in the above $k = \beta_{0k}/a$.
The $k$th zero of the Bessel function of the first kind, denoted as $J_0$, is represented by $\beta_{0k}$. It is noteworthy that in the aforementioned equation, the value of $k$ is equivalent to the ratio of $\beta_{0k}/a$.
Ultimately, the distribution of transverse energy density is expressed as follows:
\begin{equation}
\label{31}
\epsilon_{x_\perp}^{(1)}(x_\perp,\tau)=\sum _k \Bigg[\frac{2}{a^2(J_1(\beta_{0k}))^2 J_{\frac{2}{3}}(\beta_{0k}\frac{\tau_0}{a \sqrt{3}})}\int_0^a x_\perp (\epsilon_g(x_\perp,\tau_0)-\epsilon_0(\tau_0)) J_0(\beta_{1k}\frac{x_\perp}{a}) dx_\perp \Bigg] J_0(k x_\perp) \tau
^{-2/3} J_{\frac{2}{3}}(\frac{k \tau }{\sqrt{3}})
\end{equation}
\subsection{Longitudinal Expansion}

In this section, we shall examine the evolution of longitudinal expansion in an ideal fluid. The Bjorken model represents the most elementary approach to characterizing the longitudinal expansion of a fluid \cite{a1new}. 
 The Bjorken model depicts a flow that remains invariant under a Lorentz boost along the longitudinal direction. However, in reality, the longitudinal expansion may be influenced by acceleration, and non-boost invariant initial conditions may exist, leading to the absence of a rapidity plateau \cite{a9,a10,a11}. Further references and information can be found in \cite{a12,aa12,a13,a14}.

The present study investigates a model that exhibits partial breakdown of boost invariance in the longitudinal expansion. The approach taken involves the utilization of power series expansions up to the first order in $\lambda_1$, leading to the following assumption:
\begin{equation}
\label{35a}
 Y^{(1)}(\tau,\eta)=Y(\tau,\eta)-\eta.
\end{equation}
$Y^{(1)}$ denotes the acceleration of the fluid in the longitudinal direction. By combining equations (\ref{21}) and (\ref{22}), a partial differential equation for $Y^{(1)}$ can be derived, which is expressed as follows:

\begin{equation}
\label{35}
\partial_\eta^2 Y^{(1)}-3 \tau^2 \partial_\tau^2  Y^{(1)}-5 \tau \partial_\tau  Y^{(1)}=0.
\end{equation}
The equation can be solved through the method of separation of variables. The general solution is given by:
\begin{equation}
\label{36}
 Y^{(1)}(\tau , \eta)=\frac{A_0}{\tau^{2/3}}+\sum_{m=1}A_m \tau^{-\frac{1}{3}-\frac{1}{3}\sqrt{3+\frac{1}{m^2}}m}[sinh(m\eta)+B_m cosh(m \eta)].
\end{equation}
In the context,  the constant coefficients $A_0$, $A_m$, and $B_m$   can be determined upon the physical conditions. Assuming that the limit of $Y(\tau, \eta\to 0)  \to \eta$ , it can be deduced that both $A_0$ and $B_m$ must be equivalent to zero. As a result, the formula for the correction fluid rapidity $Y^1(\tau,\eta)$ can be derived.

\begin{equation}
\label{36a}
Y^{(1)}(\tau , \eta)=\sum_{m=1}A_m \tau^{-\frac{1}{3}-\frac{1}{3}\sqrt{3+\frac{1}{m^2}}m}[sinh(m\eta)].
\end{equation}
To ascertain the coefficients $A_m$, it is necessary to possess knowledge of the flow rapidity profile $Y(\tau=\tau_0, \eta)=\eta + Y^1(\tau_0,\eta)$. However, for the sake of simplicity, we limit ourselves to retaining solely the first two terms in Eq.~(\ref{36a}). Consequently, the corrected flow rapidity can be expressed as:
\begin{equation}
\label{37}
 Y^{(1)}(\tau , \eta)=\frac{A_1}{\tau} sinh(\eta)+A_2\, \tau^{-\frac{1}{3}(1+\sqrt{13})}sinh(2\eta).
 \end{equation}

Furthermore, in order to determine the correction for energy density distribution in the longitudinal direction $\epsilon^1_\eta$, it is feasible to integrate Eqs~ (\ref{21}) and (\ref{22}). As a result, the partial differential equation for $\epsilon^1_\eta$ can be formulated as follows:
\begin{equation}
\label{38}
\partial_\eta^3 \epsilon_\eta^{(1)}-3 \tau^2 \partial_\tau^2 \partial_\eta  \epsilon_\eta^{(1)}-13 \tau \partial_\tau \partial_\eta  \epsilon_\eta^{(1)}-8 \partial_\eta  \epsilon_\eta^{(1)}.
\end{equation}
The aforementioned equation can be solved through the process of separating variables. The general  solution is given by:
\begin{equation}
\label{39}
	\begin{aligned}
		& \epsilon_\eta^{(1)}(\tau , \eta)=(A_0^1+A_0^2 \eta+A_0^3 \eta^2)(\frac{A_0^{\prime 1}}{\tau^2}+\frac{A_0^{\prime 2}}{\tau^{4/3}})+(B_1^1 cosh(\eta)+ B_1^2 sinh(\eta)+B_1^3)(\frac{B_1^{\prime 1}}{\tau^{7/3}}+\frac{B_1^{\prime 2}}{\tau}) \\
		&+(C_2^1 cosh(2 \eta) + C_2^2 sinh(2 \eta)+C_2^3)(C_2^{\prime 1}\tau^{-\frac{5}{3}-\frac{\sqrt{13}}{3}}+C_2^{\prime 2}\tau^{-\frac{5}{3}+\frac{\sqrt{13}}{3}}) \\
		&+\sum_{m=3}(D_m^1 \tau^{-\frac{5}{3}-\frac{1}{3}\sqrt{3+\frac{1}{m^2}}m})(cosh(m\eta)-D_m^{\prime 1}sinh(m\eta)+D_m^{\prime 2}).
	\end{aligned}
\end{equation}
The determination of constant coefficients can be accomplished by means of physical conditions. Furthermore, by substituting both solutions, as given in Eqs.~$(\ref{39})$ and $(\ref{36a})$, into Eq.~$(\ref{22})$, one can obtain:
 \begin{equation}
 	\label{39a}
 	\begin{aligned}
 		& \epsilon_\eta^{(1)}(\tau , \eta)=(B_1^1 cosh(\eta))(\frac{B_1^{\prime 1}}{\tau^{7/3}}+\frac{B_1^{\prime 2}}{\tau}) \\
 		&+(C_2^1 cosh(2 \eta) )(C_2^{\prime 1}\tau^{-\frac{5}{3}-\frac{\sqrt{13}}{3}}+C_2^{\prime 2}\tau^{-\frac{5}{3}+\frac{\sqrt{13}}{3}}) \\
 		&+\sum_{m=3}(D_m^1 \tau^{-\frac{5}{3}-\frac{1}{3}\sqrt{3+\frac{1}{m^2}}m})(cosh(m\eta)).
 	\end{aligned}
 \end{equation}

Upon consideration of the particular solution represented by Eq.~(\ref{37}) pertaining to the fluid rapidity, the solution for the correction energy density $\epsilon^1_\eta$ can be expressed as follows:
\begin{equation}
\label{40}
 \epsilon_\eta^{(1)}(\tau , \eta)= \frac{4}{3} \epsilon_c \tau_0^{4/3} A_1 \frac{ cosh(\eta)}{\tau^{7/3}}+ \frac{2}{3}(\sqrt{13}-1) \epsilon_c \tau_0^{4/3} A_2\tau^{-\frac{5}{3}-\frac{\sqrt{13}}{3}} cosh(2 \eta).
\end{equation}
The above equation comprises of two distinct terms that delineate the energy density distribution in the longitudinal expansion. The following section will elucidate the behavior of these physical quantities.

\section{Results and discussions}
In this section, we provide a comprehensive analysis of the dynamical evolution and characteristics of our model, utilizing a perturbation approach. Our investigation focuses on the longitudinal and radial evolution of the fluid, which is reflected in the transverse fluid velocity and acceleration parameter, respectively, due to the influence of the generalized Bjorken model. Through our perturbation approach, we derive the corrected fluid velocities, acceleration parameter, and energy density. These quantities provide valuable insights into the space-time evolution of the quark-gluon plasma in heavy-ion collisions.

The aforementioned quantities are crucial in comprehending the space-time evolution of the quark-gluon plasma in heavy-ion collisions. To accurately assess these quantities, it is imperative to establish the values of the constants $A_1$, $A_2$, $q$, and $\hat{\epsilon_0}$. These  are the only  free parameters  in our model.

The initial conditions for the transverse expansion of the Quark-Gluon Plasma (QGP) are characterized by two parameters, namely $q$ and $\hat{\epsilon_0}$, which are introduced by Gubser's solution.
 Together,  they determine the initial energy density  profile  of the plasma at some  early time that should be comparable
to or greater than the time at which a hydrodynamic
description becomes valid. The
parameter $q$   also implicitly determines the radial
velocity profile at  early time of the hydrodynamic evolution. Our approach, therefore, is to explore the two
parameter space looking for  reasonable
values to mock up heavy ion collisions.  We have found that
choosing $\hat{\epsilon_0}=1500$  and $1/q=6.4 \ fm$ yields reasonable results, as we shall show below.

In order to accurately depict the space-time assessment of longitudinal expansion of QGP, it is imperative to establish fixed parameters, namely $A_1$ and $A_2$, that align with phenomenological analyses. Regrettably, in this particular study, the aforementioned parameters have been selected based on the following condition:
\begin{equation}
\label{40a}
\frac{  \epsilon_\eta^{(1)}(\tau , \eta)}{\epsilon_0(\tau)} < \frac{  \epsilon_\eta^{(1)}(\tau_0 , \eta=0)}{\epsilon_0(\tau_0)}=\frac{4}{3} A_1 \frac{1}{\tau_0}+ \frac{2}{3}(\sqrt{13}-1) A_2\tau_0^{-\frac{1}{3}-\frac{\sqrt{13}}{3}} <<1.
\end{equation}
 For our numerical computation, we have opted to utilize the values of $A_1=0.3$ and $A_2=-0.07$.

\subsection{ Transverse Expansion}
This subsection presents the numerical results of the transverse velocity and energy density obtained through our perturbation approach. These two quantities aid in comprehending the transverse evolution of the quark-gluon plasma in heavy ion collisions. Through our analysis, the value of $\epsilon_0$ has been determined to be $5.4 GeV/fm^3$ at a proper time of approximately $\tau = 1\  fm$, as reported in Ref.~\cite{a33}. The parameters $\hat{\epsilon_0}=1500$ and $1/q=6.4$ fm have been chosen for our study. 
Subsequently, we display  the transverse fluid velocity $(v_{\perp}=\frac{u_\perp}{u_\tau} )$ and the transverse energy density distribution ($\epsilon^1_\perp({\tau,x_\perp})$).

\begin{figure}[H]
   	\centering
   	{\includegraphics[width=.45\textwidth]{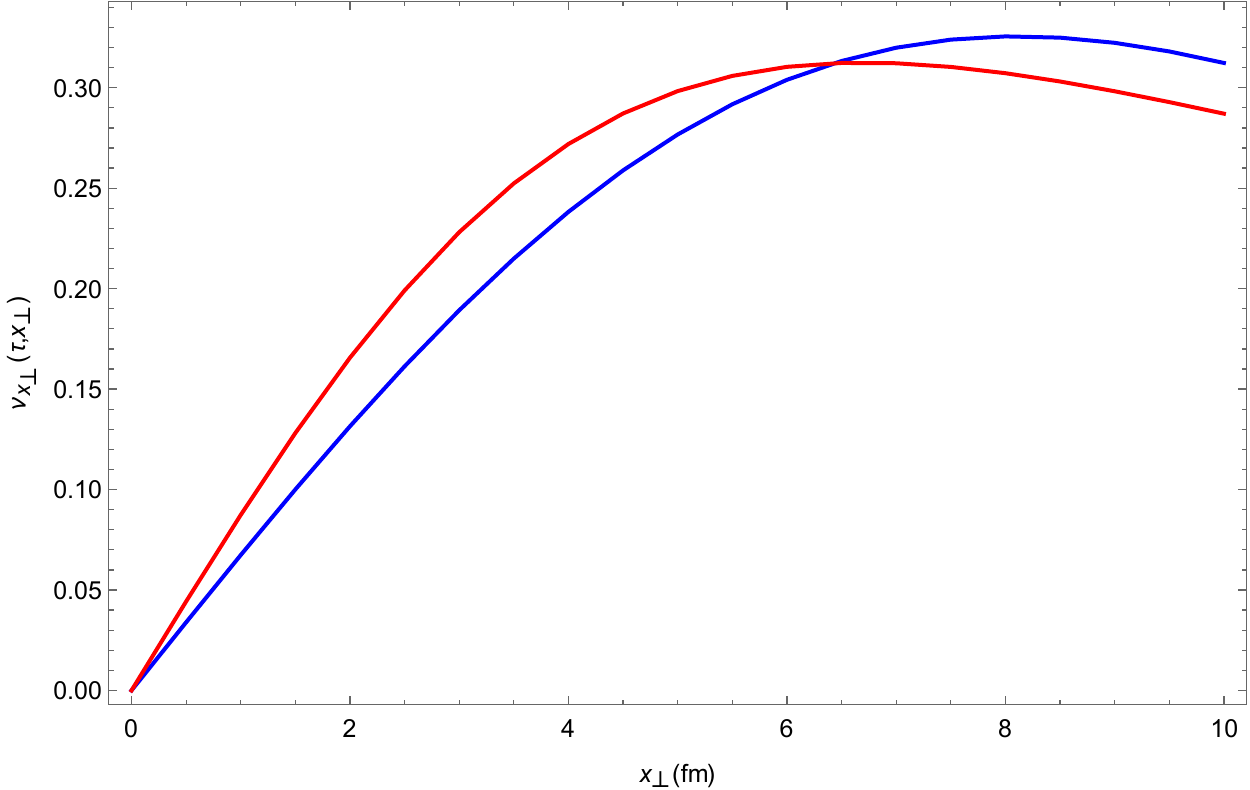}}
   	{\includegraphics[width=.45\textwidth]{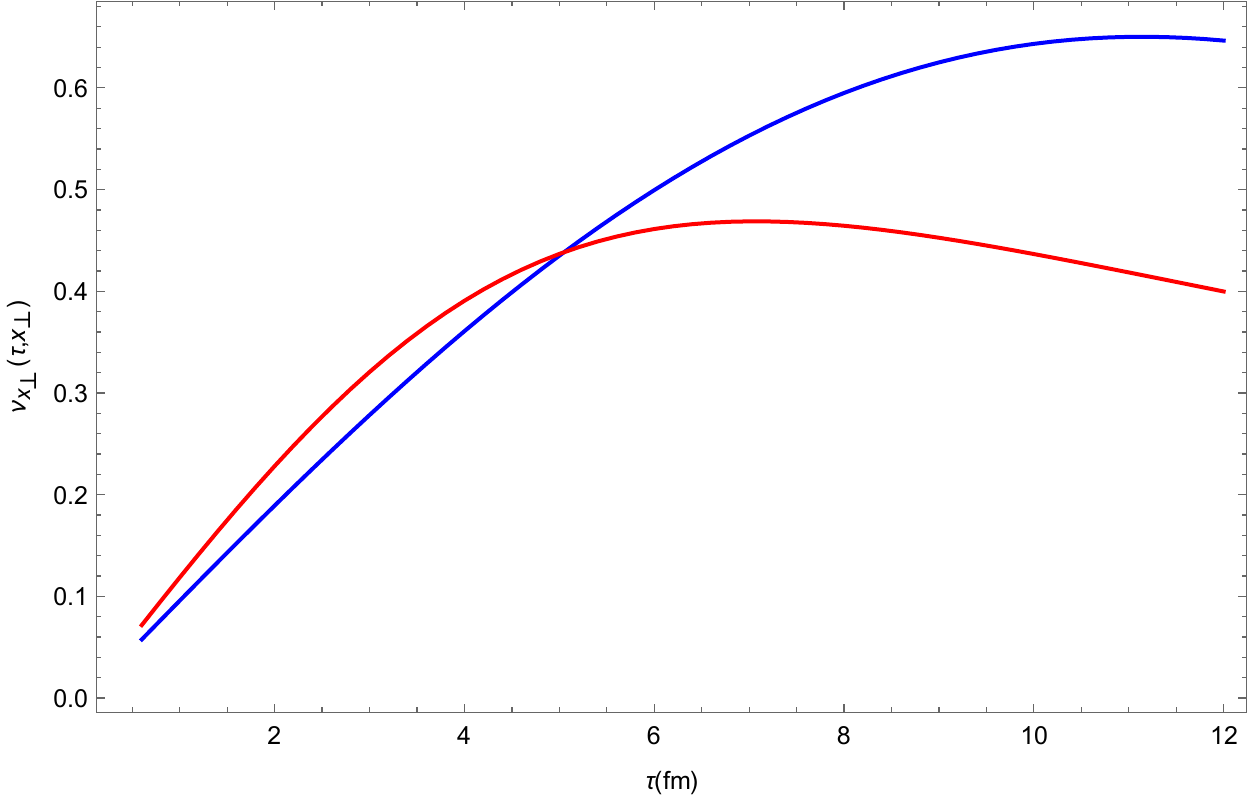}}
   	\caption{\small{ Left: The transverse velocity  $v_\perp=\frac{u^\perp}{u^\tau}\simeq K^1(\tau,x_\perp)$ in terms of  $x_\perp$ for  $\tau=2$. Right: The  transverse velocity  in terms of $\tau$ for $x_\perp=3$. The blue curve correspond to present work and the red curve correspond to \cite {a35} }. It is  measured in $fm/c$, with  $q=1/6.4(fm)^{-1}$. 
   		\label{f1}}
   \end{figure}

\begin{figure}[H]
   	\centering
   	{\includegraphics[width=.45\textwidth]{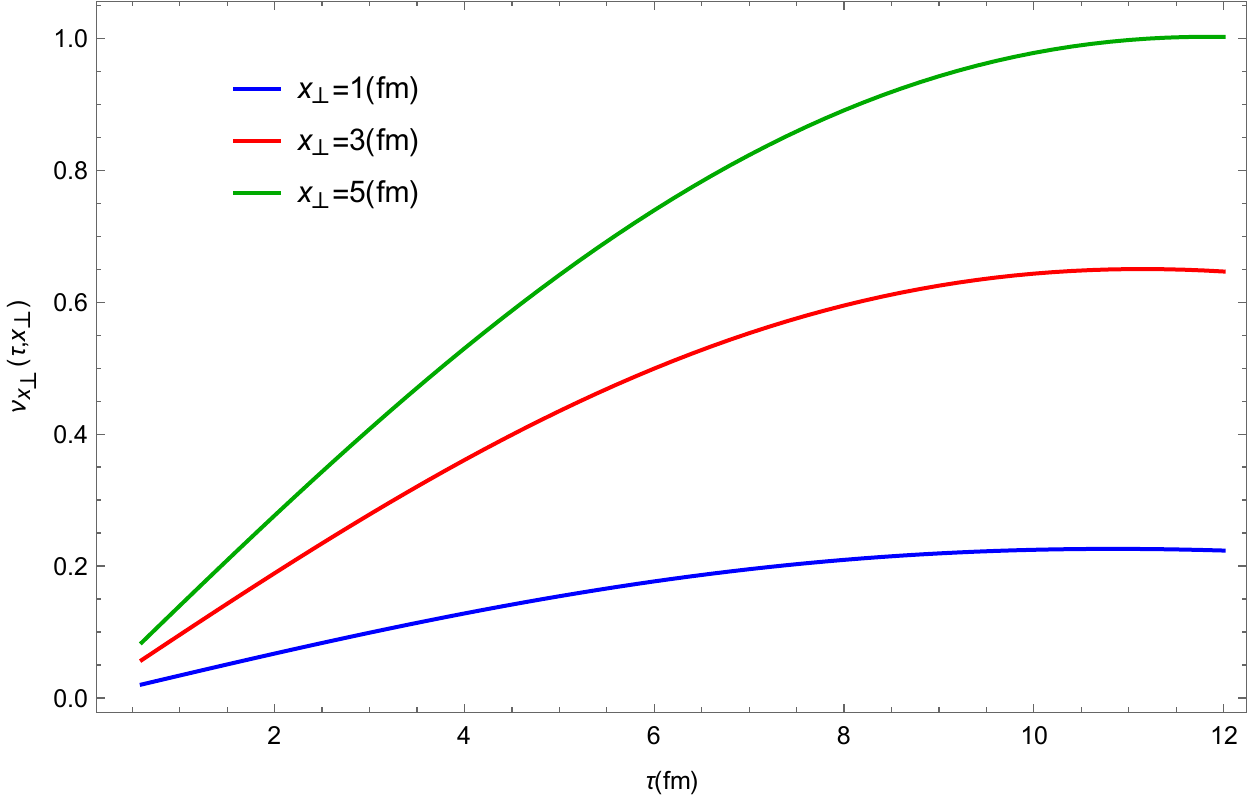}}
   	{\includegraphics[width=.45\textwidth]{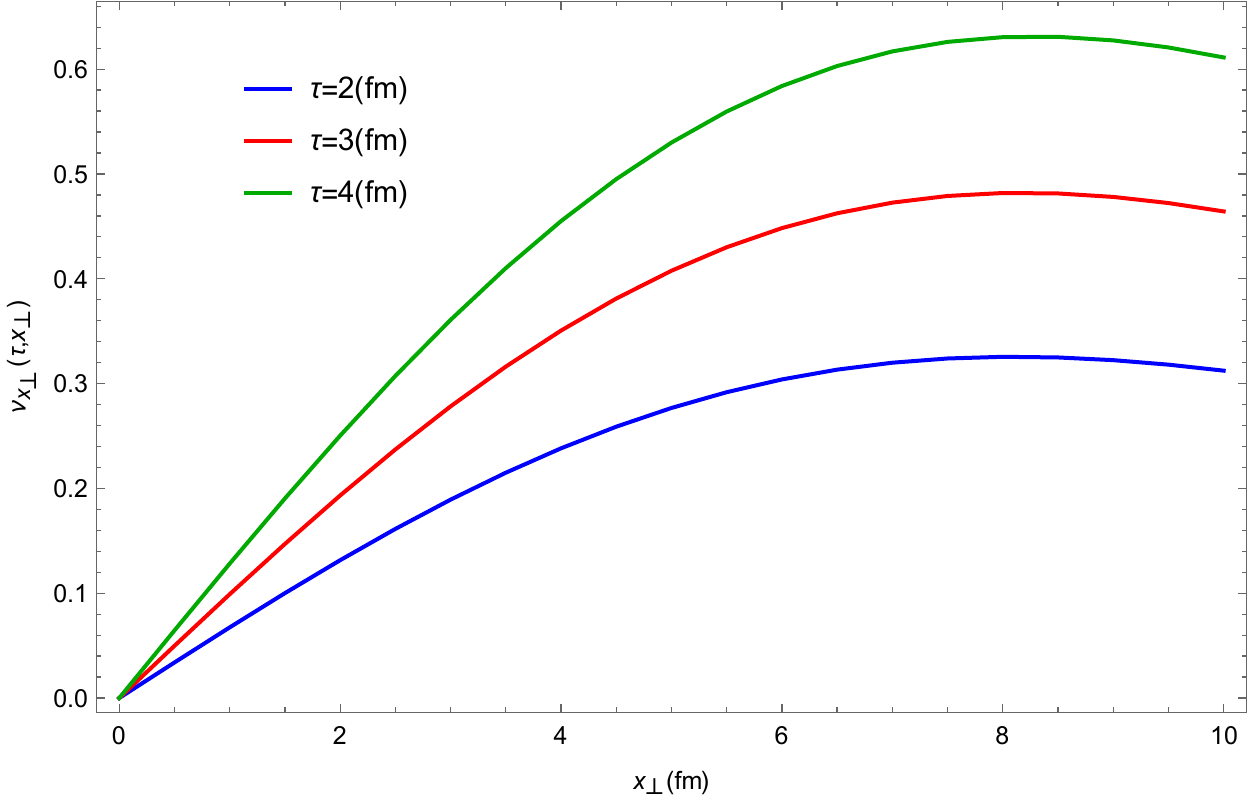}}
   	\caption{\small{ Left: The transverse velocity $v_\perp$
   			as a function of proper time $\tau$ for several values of  transverse radius $x_\perp$. Right: The transverse velocity $v_\perp$
   			as a function of  transverse radius $x_\perp$  for several values of  proper time $\tau$. It is  measured in $fm/c$, with  $q=1/6.4(fm)^{-1}$.  }}
   	\label{f2}
   \end{figure}

\begin{figure}[H]
   	\centering
   	{\includegraphics[width=.55\textwidth]{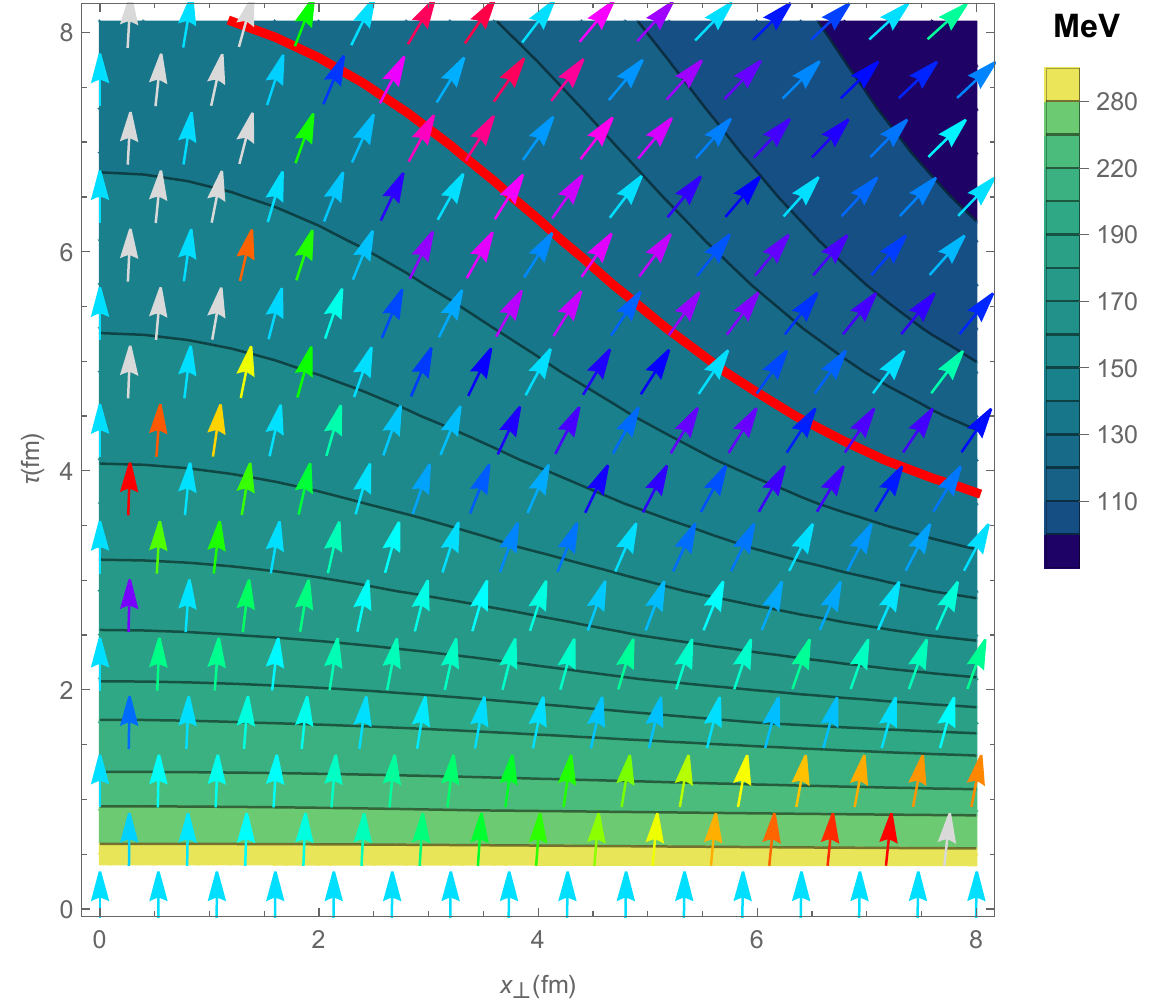}}
   	\caption{The two-dimensional  fluid velocity $(u^\tau/\sqrt{u_\tau^2+u_\perp^2},u^\perp/\sqrt{u_\tau^2+u_\perp^2})$ is plotted with parameters chosen as $q=1/6.4(fm)^{-1}$, $\tau_0=1 (fm)$, $\hat{\epsilon}_0=1500$, and $\epsilon_c=5.4$ ${GeV}/{fm^3}$ }
   	\label{f2a}
   \end{figure}

\begin{figure}[H]
   	\centering
   	{\includegraphics[width=.45\textwidth]{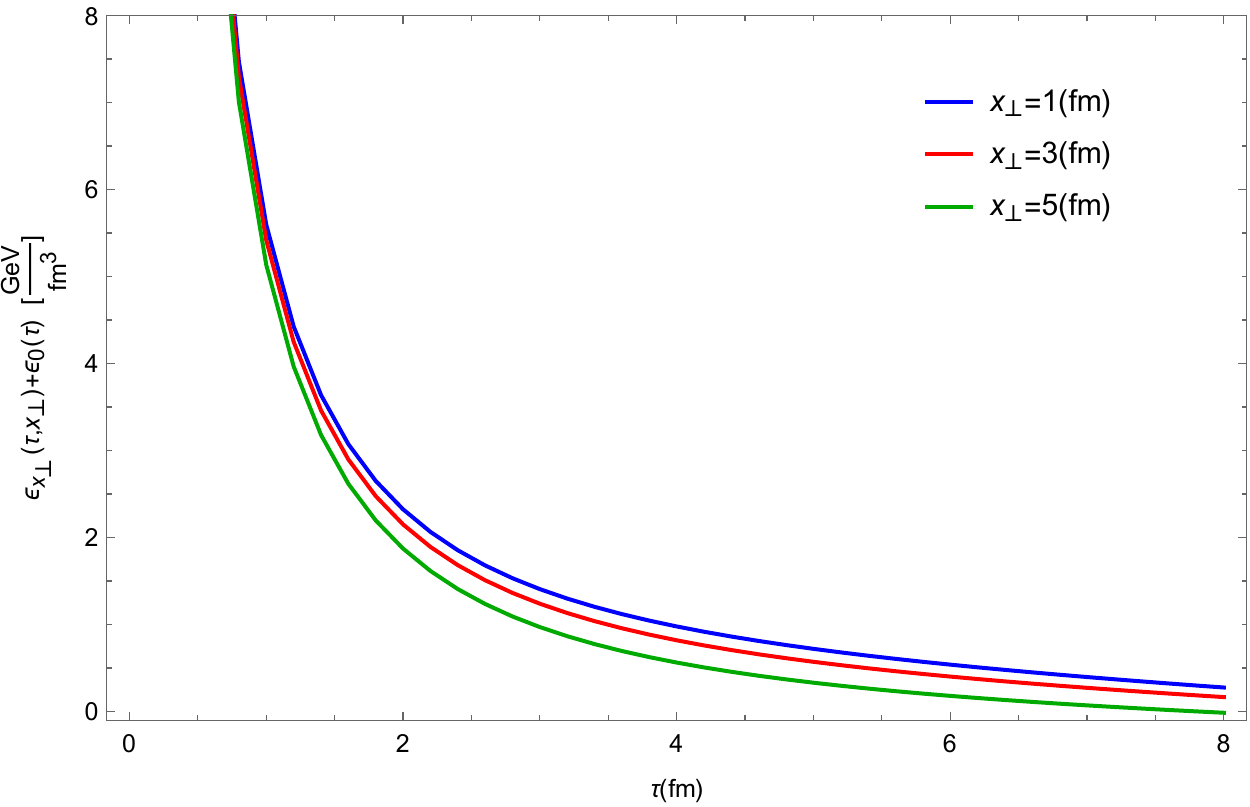}}
   	{\includegraphics[width=.45\textwidth]{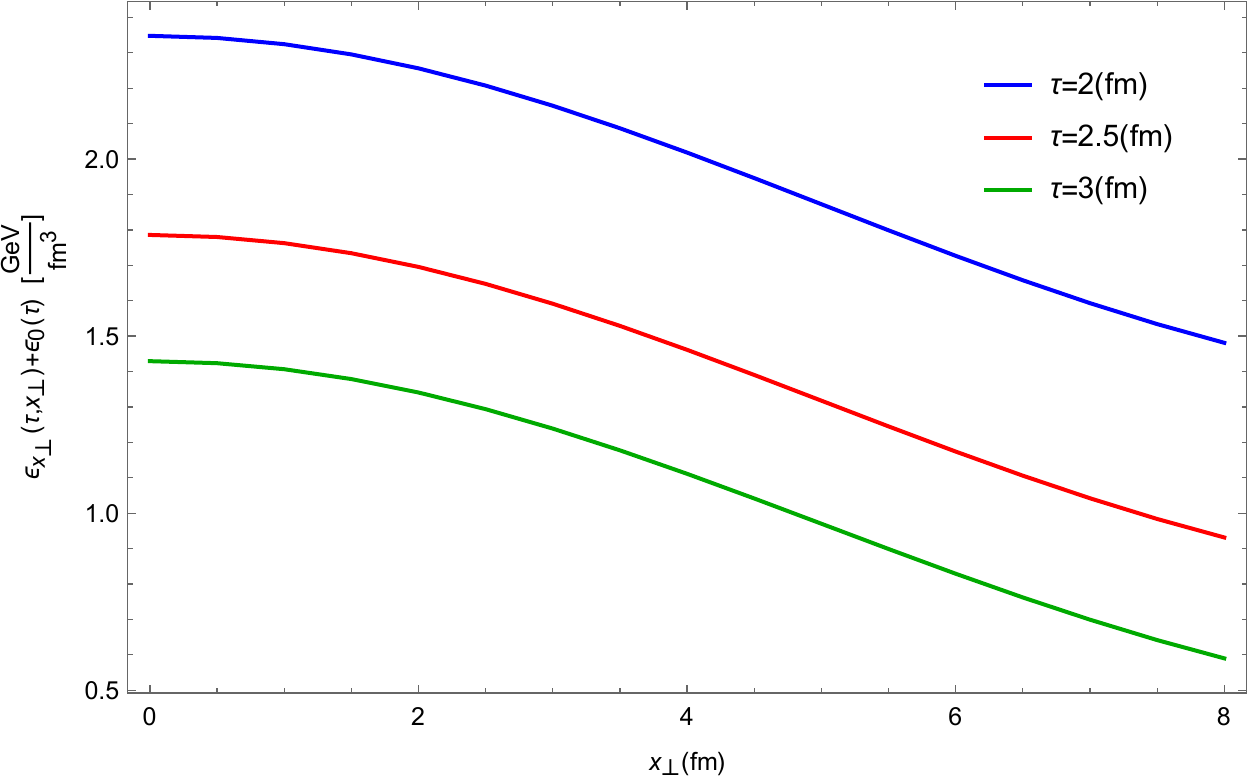}}
   	\caption{\small{Left: $\epsilon(\tau,x_\perp)$ in terms of      proper time  $\tau$  for different value of  $x_\perp$. Right: $\epsilon(\tau,x_\perp)$ in terms of    transverse radius  $x_\perp$      for different value of $\tau$. With parameters chosen as $q=1/6.4(fm)^{-1}$, $\tau_0=1 (fm)$, $\hat{\epsilon}_0=1500$, and $\epsilon_c=5.4$ ${GeV}/{fm^3}$.	  }}
   	\label{f4}
   \end{figure}

Fig.~(\ref{f1}) illustrates the transverse velocity, denoted as $v_\perp$, which is defined as the ratio of the transverse component of the four-velocity, $u^\perp$, to its temporal component, $u^\tau$. The plot depicts $v_\perp$ as a function of $x_\perp$ or $\tau$, while either $\tau$ or $x_\perp$ is held constant. Additionally, a comparison with Gubser's transverse velocity is presented.
 The transverse velocity's dependence on $x_\perp$ exhibits a comparable shape to that of Gubser's work. Nevertheless, the transverse velocity's dependence as a function of $\tau$, as depicted in the right panel of Fig.~(\ref{f1}), deviates from the Gubser flow for $\tau>5 \ fm$. This suggests that our flow exhibits a longer life time compared to the Gubser flow. Fig.~(\ref{f2}) displays the transverse velocity $v_\perp$ as a function of either $\tau$ or $x_\perp$ for various values of $x_\perp$ or $\tau$, respectively. Notably, it is observed that $v_\perp$ increases for larger values of $\tau$ (at a fixed $x_\perp$) or for larger values of $x_\perp$ (at a fixed $\tau$). This phenomenon is a consequence of the assumption of conformal symmetry in the initial conditions, which impacts the spatiotemporal evolution of the transverse expansion of the fluid \cite{a35}.  Nevertheless, the energy density experiences a significant decrease and ultimately reaches zero at a  large value of $x_\perp$, as depicted in the right panel of Fig.~(\ref{f4}). This peculiar phenomenon is attributed to a complete failure of the derivative expansion, which serves as the foundation of hydrodynamics. To elucidate this aspect, we present the two-dimensional fluid velocity and the contours of constant temperature in Fig.~(\ref{f2a}). The plot has been exhibited for values of $(\tau,x_\perp)$ wherein $\epsilon$ is positive. The prominently delineated red contour corresponds to the temperature of $130\ MeV$.  It is suggested that the region of flow in which the temperature exceeds 130 MeV may be considered as the quark-gluon plasma (QGP). It is important to note that the temperature of 130 MeV is approximately the decoupling temperature in a Cooper-Frye treatment. Beyond this temperature, the genuine degrees of freedom are those of a nearly free hadron gas, rather than a fluid.
 The hydrodynamic approximation is deemed invalid when the temperature falls below 130 MeV. In summary, our hydrodynamic model approximation is applicable only for values of $x_\perp$ and $\tau$ where  the temperature exceeds 130 MeV.  A comparison with the findings of reference \cite{a35} suggests that a slightly larger value of $\tau$ and a smaller value of $x_\perp$ would be more appropriate in our case. This indicates that the spatial regain  of the QGP is smaller compared to the Gubser flow, but the  lifetime of the system in the plasma phase is longer than in the Gubser case.

\begin{figure}[H]
   	\centering
   	{\includegraphics[width=.45\textwidth]{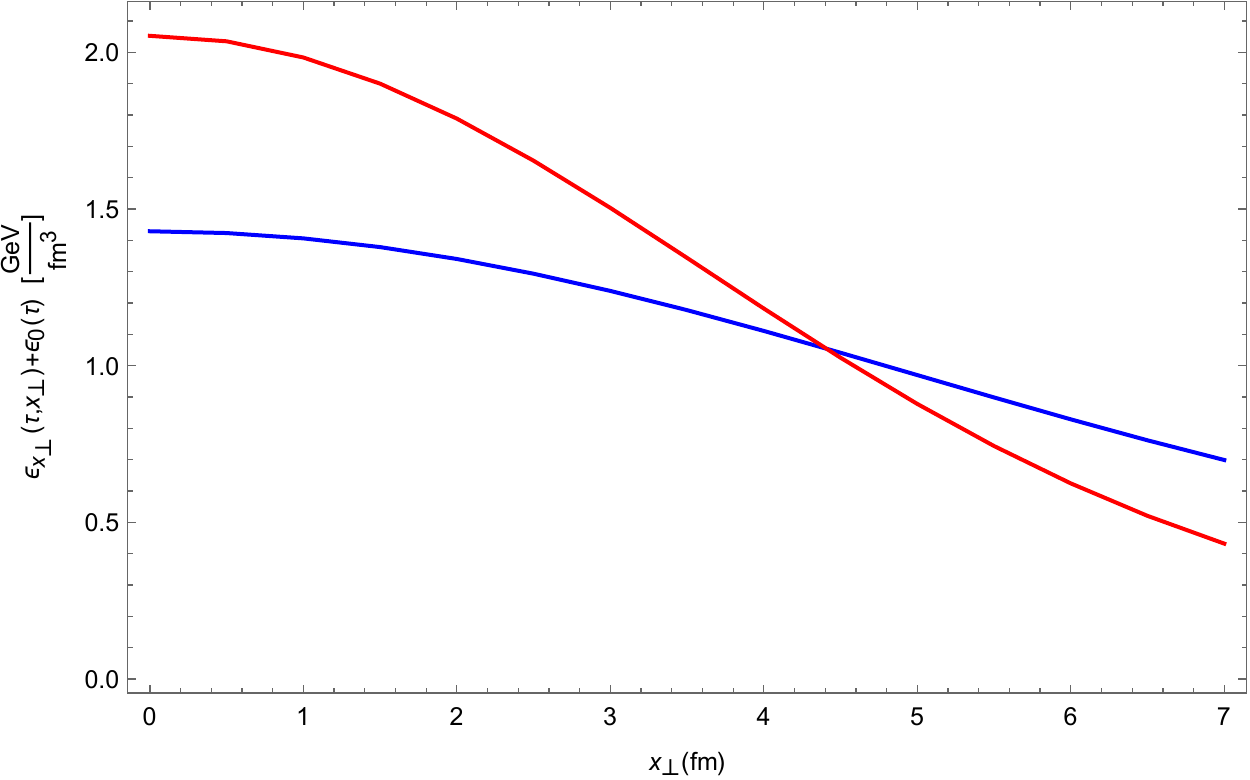}}
   	{\includegraphics[width=.45\textwidth]{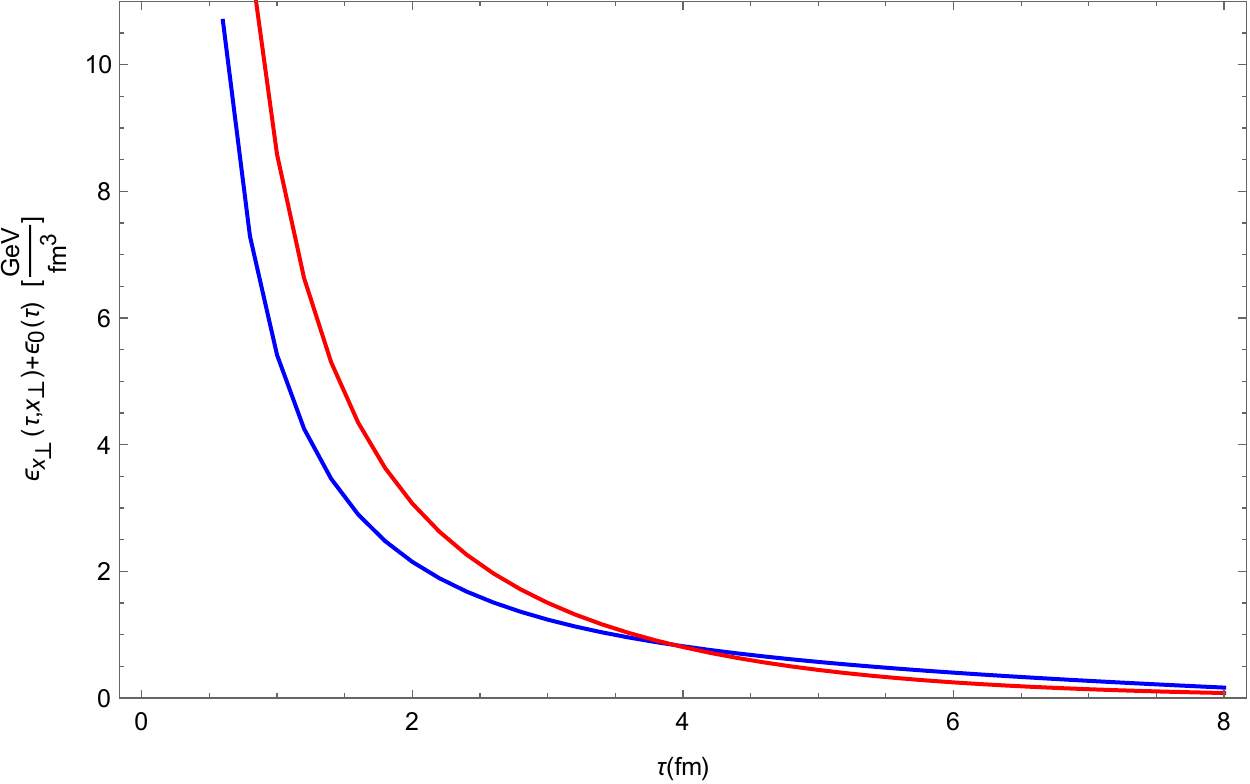}}
   	\caption{\small{ Left: 
   			The total energy density 
   			$\epsilon(\tau,x_\perp)$ 
   			as a 	 function of transverse radius  $x_\perp$  at $\tau=3 \ fm$. Right:  The total energy density 
   			$\epsilon(\tau,x_\perp)$ 
   			as a 	 function of proper time  $\tau$  at $  x_\perp=3 \ fm$. The blue curve correspond to present work and the red curve  to \cite {a35}. With parameters chosen as $q=1/6.4(fm)^{-1}$, $\tau_0=1 (fm)$, $\hat{\epsilon}_0=1500$, and $\epsilon_c=5.4$ ${GeV}/{fm^3}$. }}
   	\label{f3}
   \end{figure}

   Fig.~(\ref{f3}) displays the energy density $\epsilon(\tau,x_\perp)=\epsilon_0(\tau) +\epsilon^1_{x_\perp}(\tau,x_\perp)$ as a function of $x_\perp$ or $\tau$ at a fixed value of either $\tau$ or $x_\perp$. It is important to note that the total energy density is given by $\epsilon(\tau,x_\perp,\eta)=\epsilon_0+\epsilon^1_{x_\perp}+\epsilon^1_\eta$. The analysis also includes a comparison with the Gubser flow. The results indicate that the majority of energy distribution is concentrated at small values of $x_\perp<5 \ fm$. 
   Furthermore, the left panel of Fig.~(\ref{f3}) demonstrates that the spatial distribution of energy density is comparatively smoother than that of the Qubser flow's energy density.
        Fig.~(\ref{f4})  shows  the  energy density $\epsilon(\tau,x_\perp)/\epsilon_0$  as a function of  $x_\perp$  for    different values of $\tau$ or as a function of $\tau$ for different values of $x_\perp$, respectively.

   To evaluate the efficacy of our model, we derive the coefficients $c^k$ in (\ref{25}) by employing an alternative radial flow profile, as investigated in \cite{a25}, at the initial proper time $\tau_0$. The coefficients $c^k$ are obtained through the utilization of the orthogonality of Bessel functions, and are expressed as follows:
   \begin{equation}
   \label{28f}
   c^k=\frac{2}{a^2(J_2(\beta_{1k}))^2 J_{\frac{1}{3}}(\beta_{1k}\frac{\tau_0}{a \sqrt{3}})}\int_0^a \frac{tanh( x_\perp/50)}{\bar{\gamma}} J_1(\beta_{1k}\frac{x_\perp}{a}) dx.
   \end{equation}
  Upon conducting a first order expansion calculation, it has been determined that $\bar{\gamma}\simeq 1$. As a result, the transverse fluid velocity can be expressed as:
   \begin{equation}
   \label{25aa}
  v^\perp \simeq  u^{(1)}(x_\perp,\tau)= \sum _k\Bigg[\frac{2}{a^2(J_2(\beta_{1k}))^2 J_{\frac{1}{3}}(\beta_{1k}\frac{\tau_0}{a \sqrt{3}})}\int_0^a  tanh( x_\perp/50)  J_1(\beta_{1k}\frac{x_\perp}{a}) dx_\perp \Bigg] J_1(k x_\perp) \tau^{2/3}  J_{\frac{1}{3}}(\frac{k \tau }{\sqrt{3}}).
   \end{equation}

\begin{figure}[H]
   	\centering
   	{\includegraphics[width=.40\textwidth]{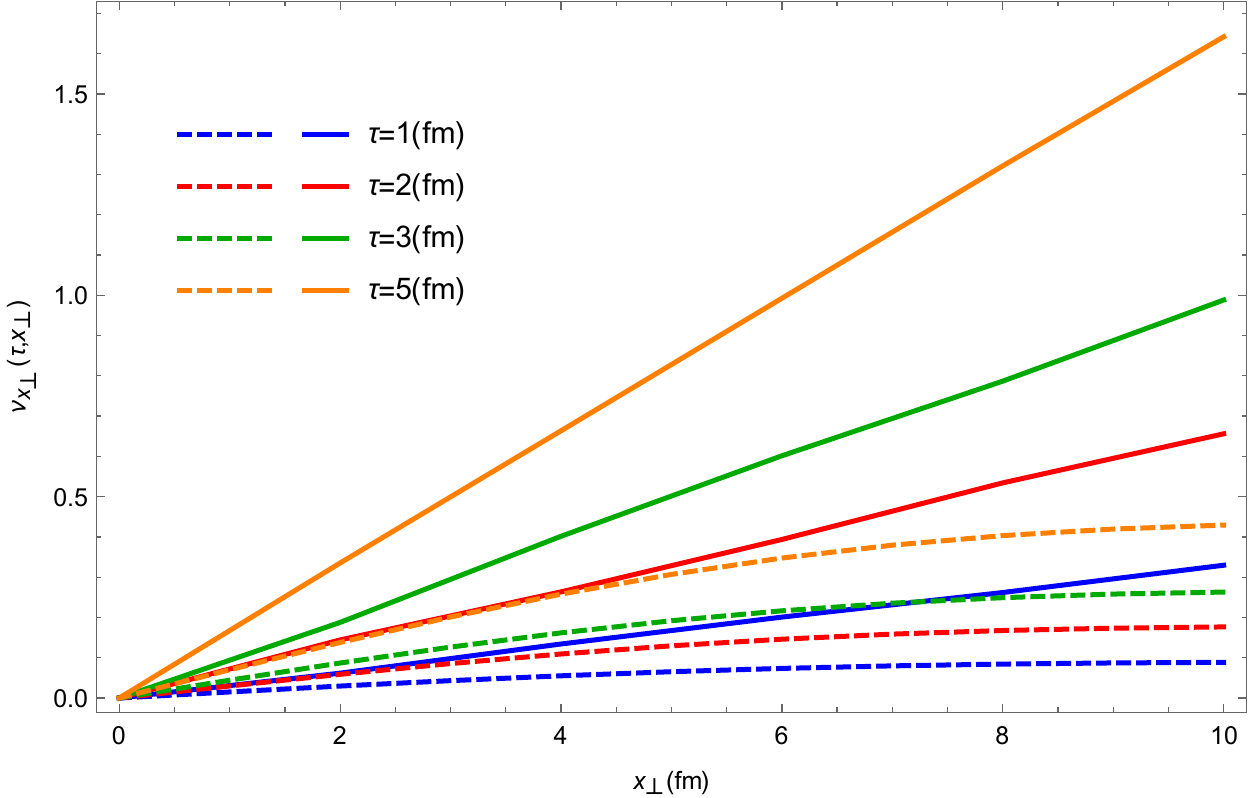}}
   	{\includegraphics[width=.40\textwidth]{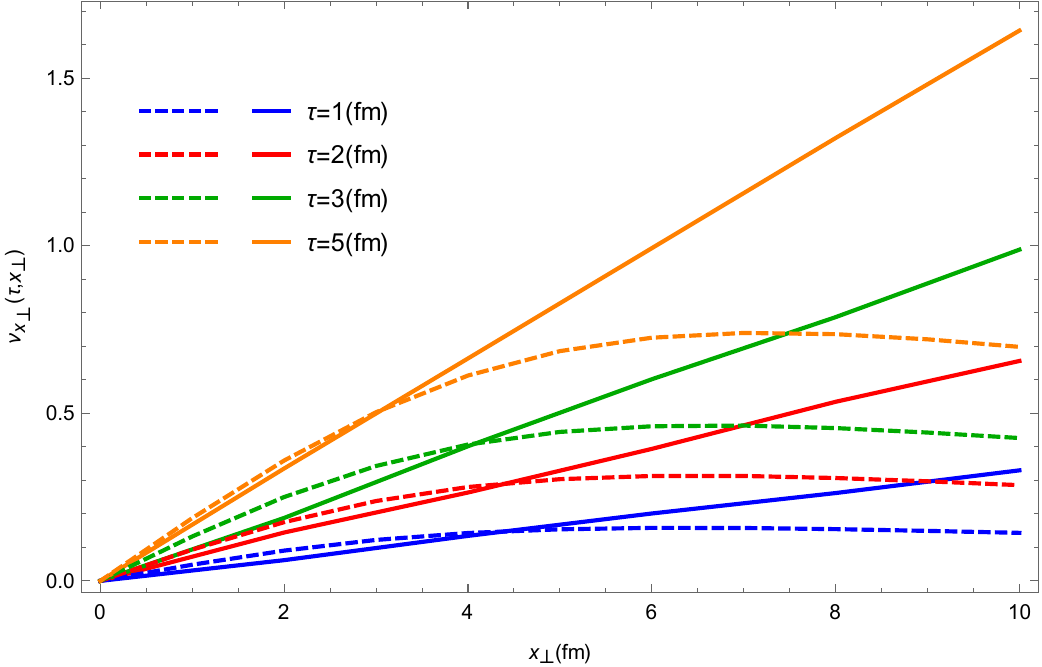}}
   	\caption{\small{ The transverse velocity $v_\perp$ as a function of transverse radius $x_\perp$ for several values of $\tau$, measured in fm. The  solid  lines displayed   the transverse velocities where $c^k$ are  obtained from   the phenomenological proposal $v_\perp = tanh x_\perp/50$ at $\tau_0=0.6$ fm. The dashed lines displayed  the transverse velocities where $c^k$ are  obtained from   the  Gubser proposal at $\tau_0=0.6$ fm. The initial proper time is chosen as $\tau_0=0.6$ fm.	Two different values of  $q=1/11.24266(fm)^{-1}$ (left panel) and $q =1/6.4(fm)^{-1}$ (right panel) are chosen for the   Gubser proposal as initial conditions. 		
   	}}
   	\label{f4a}
   \end{figure}

  Fig.~(\ref{f4a}) displays the transverse velocity $v_\perp \simeq u^1$ as a function of transverse radius $x_\perp$ for various values of the proper time $\tau$. The radial flows are compared, with the coefficients $c^k$ being derived from distinct initial conditions.  The transverse velocity was displayed by the solid lines utilizing the initial conditions derived from the phenomenological proposal $v_\perp=tanh\frac{x_\perp}{50}$ at $\tau_0=0.6$. The transverse velocity, derived from the initial conditions proposed by Gubser at $\tau_0=0.6$ for two distinct values of $q$, was represented by dashed lines in the plot. The results of our study suggest that, for a value of $q$ equal to $1/6.4 \ (fm)^{-1}$ and $x_\perp<5$, the radial velocity of systems with varying initial conditions converge to a common late-time behavior. However, it is observed that this convergence does not occur when $q$ deviates from the aforementioned value, as depicted in the left panel of Fig.(\ref{f4a}).
   This finding is in accordance with the results presented in \cite{a26}, which have also concluded that selecting $q = 1/6.4 \ fm$ produces reasonable spectra for both pions and protons.
   
   \subsection{ The Longitudinal Expansion}
 The QGP system that undergoes evolution in relativistic heavy ion collisions is inherently non-boost-invariant. The hot medium is confined within a finite range of rapidity, while the system remains dilute outside of this window. Within this subsection, we present a novel set of solutions to the hydrodynamic equations that do not exhibit boost-invariance.  
 We shall illustrate the impact of a disrupted longitudinal boost invariant motion on the solutions of relativistic hydrodynamics. Our general solutions, denoted as($\ref{36a}$) and ($\ref{39a}$), delineate the effects of non-boost longitudinal expansion of the fluid. These solutions have been employed to approximate the correction fluid rapidity $v_\eta(\tau,\eta)\simeq Y^1(\tau,\eta)$ and the correction longitudinal energy density $\epsilon^1_\eta$ that arise from the hydrodynamic solutions expressed in Eqs~$ (\ref{37}, \ref{40})$.

\begin{figure}[H]
 	\centering
 	{\includegraphics[width=.45\textwidth]{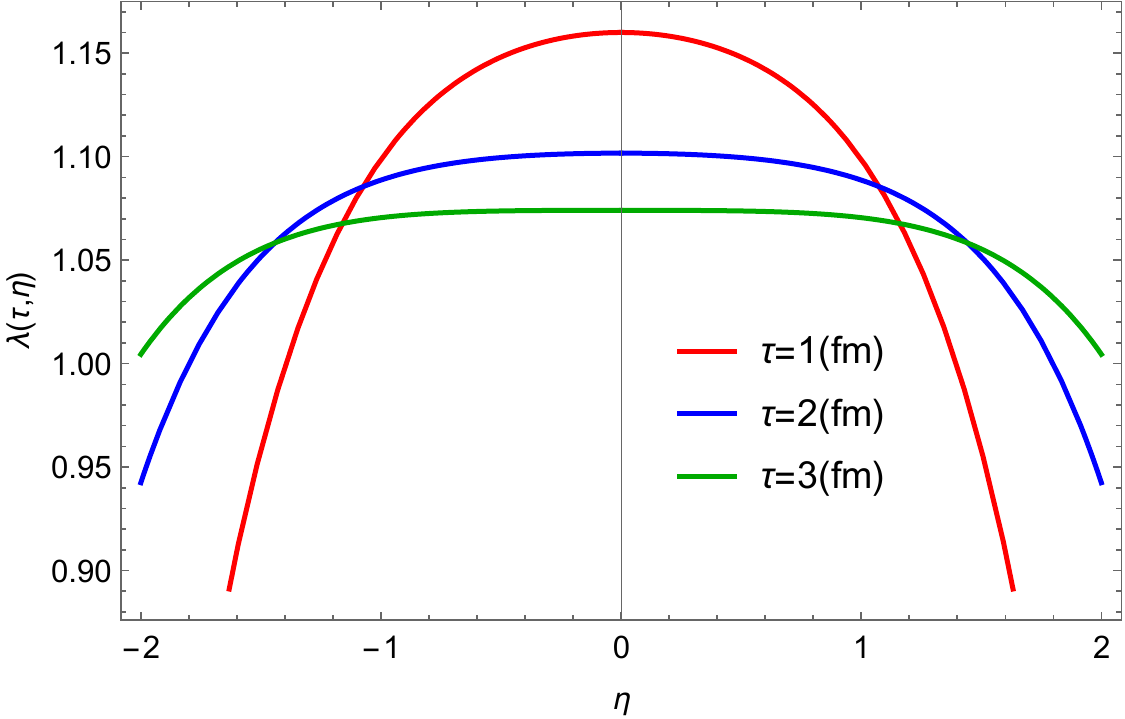}}
 	{\includegraphics[width=.45\textwidth]{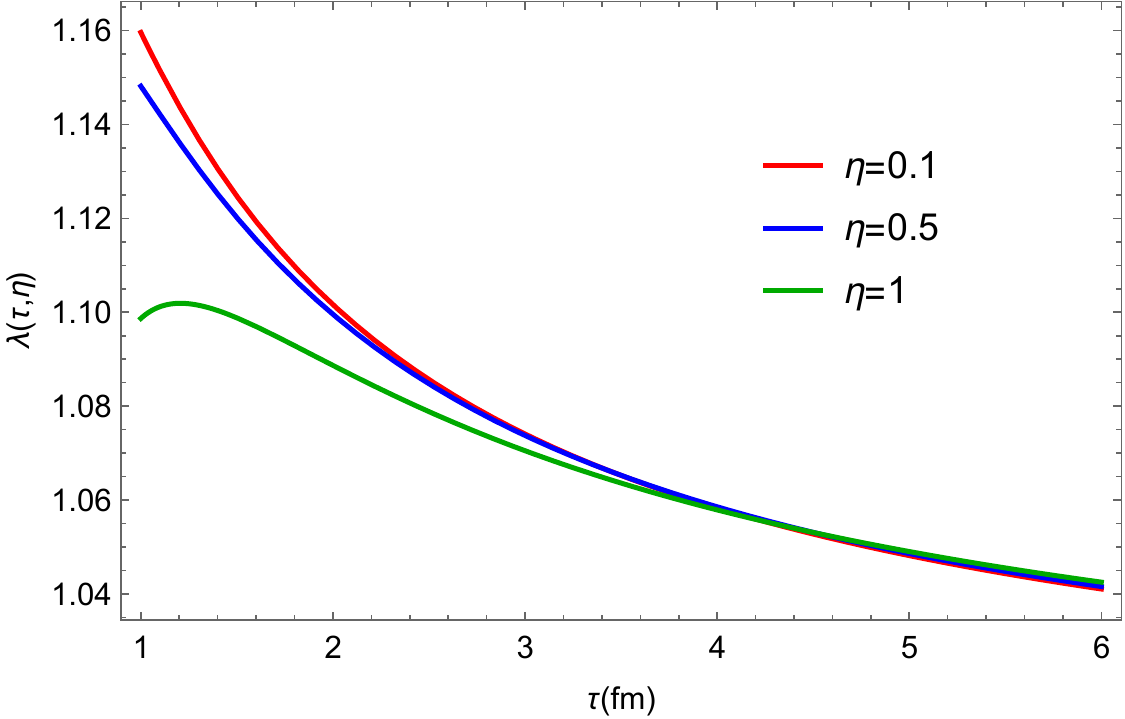}}
 	\caption{\small{Right: Acceleration parameter 	$\lambda(\tau,\eta)=Y(\tau,\eta)/ \eta \simeq \frac{Y^1(\tau,\eta)+\eta}{\eta}$  in terms of $\eta$ with different value of $\tau$. Right: 			
 			$\lambda(\tau,\eta)$ in terms of  $\eta$ with different value of $\eta$.  
 			The values  $ A_1 = 0.3$ and $A_2= -0.07$ are chosen. }}
 	\label{f5}
 \end{figure}

In this study, we have conducted an investigation into the dynamical evolution of the fluid rapidity. The results of this investigation are presented in Fig.~(\ref{f5}), which depicts the evolution of $\lambda=\frac{Y}{\eta}=\frac{Y^1(\tau,\eta)+\eta}{\eta}$ as a function of $\eta$ for various values of $\tau$ (the left panel), or as a function of $\tau$ for different values of $\eta$ (the right panel). It is important to note that $\lambda$ is commonly referred to as the acceleration parameter\cite{a13}, as it characterizes the acceleration of the longitudinal flow. 
 The  analysis reveals that the acceleration parameter exhibits a decreasing trend as the absolute value of $\eta$ increases. However, it is noteworthy that a plateau is observed for the late-time regime. The plotted data indicates that an increase in $\eta$ leads to a reduction in the acceleration parameter, and over time, the attenuation of $\lambda (\tau, \eta)$ occurs over a broad range of rapidity. This implies that the flow experiences a greater acceleration at lower values of $\eta$ or $\tau$.

\begin{figure}[H]
 	\centering
 	{\includegraphics[width=.45\textwidth]{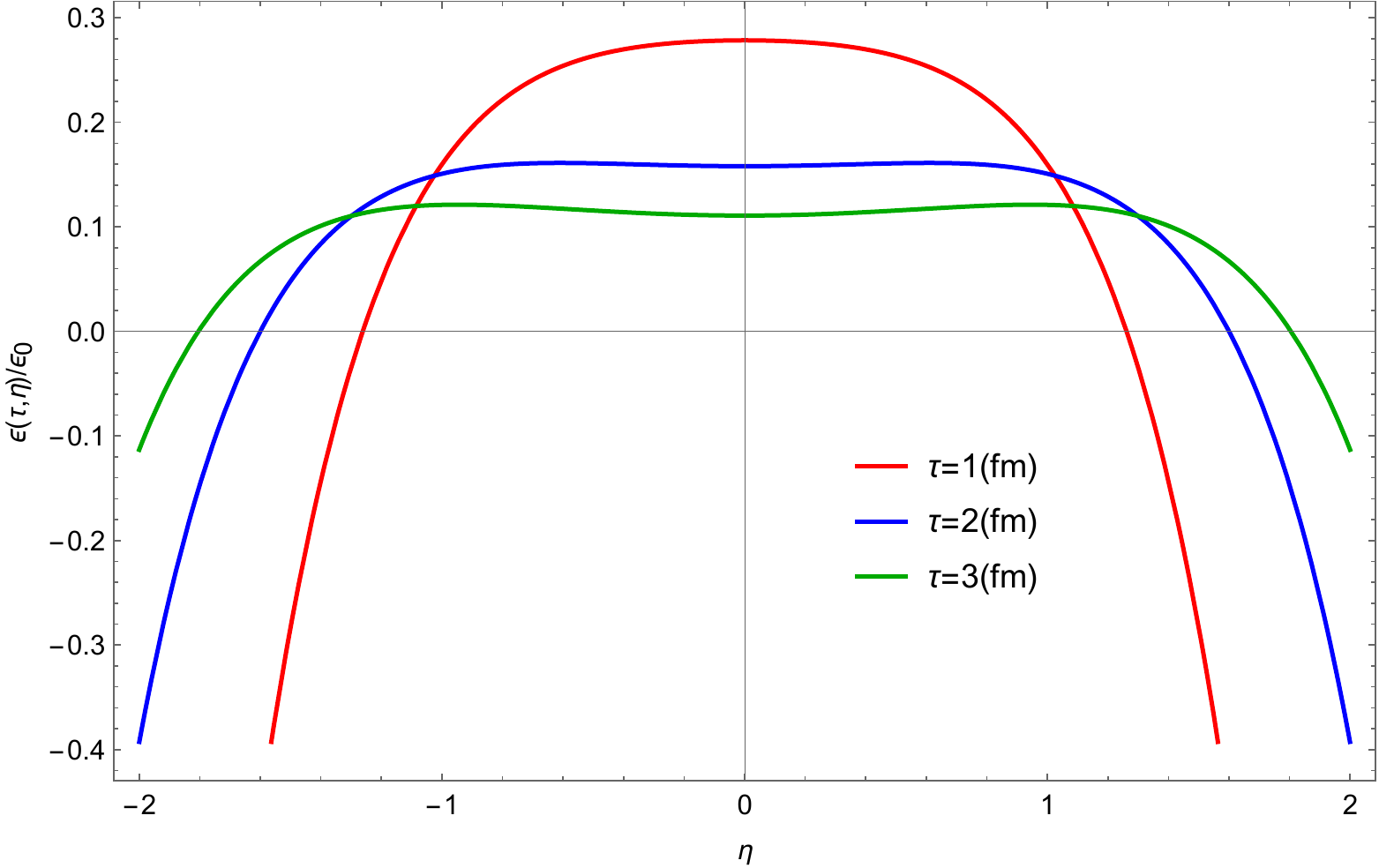}}
 	{\includegraphics[width=.45\textwidth]{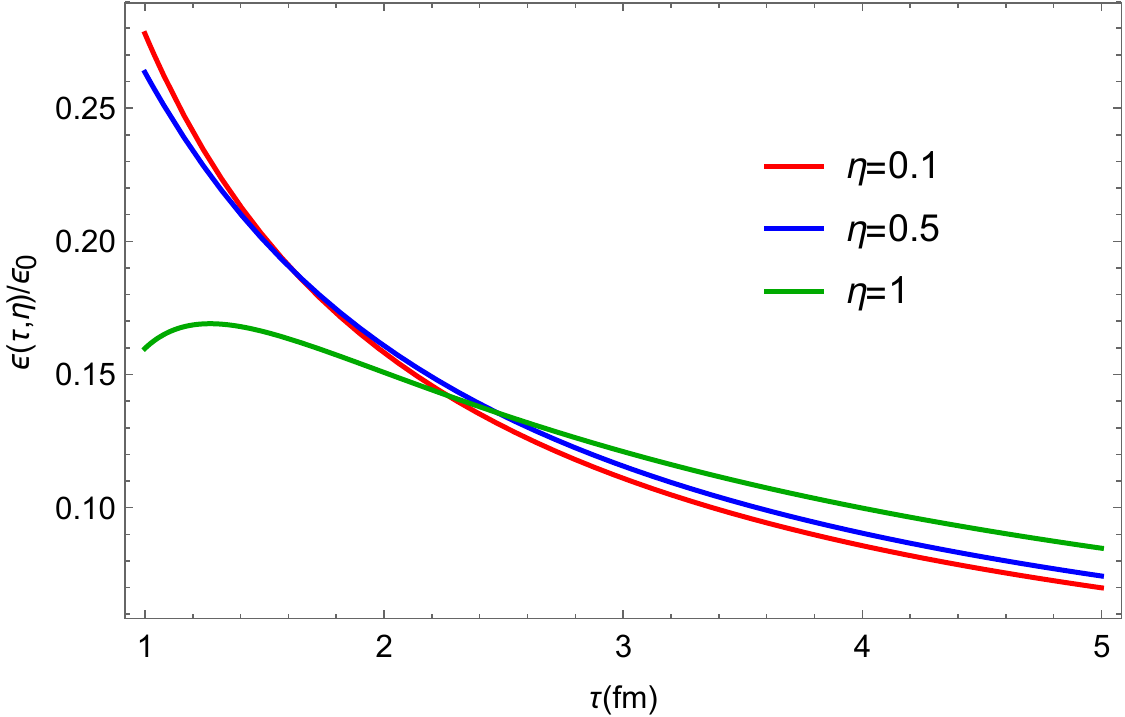}}
 	\caption{\small{Right:
 			$\epsilon_\eta(\tau,\eta)/\epsilon_0$  as a 	 function of   the space-time rapidity $\eta$ with different values of  $\tau$. Right: $\epsilon_\eta(\tau,\eta)/\epsilon_0$ as a function of proper time $\tau$ with different value of $\eta$. 
 			The values $ A_1 = 0.3$ and $A_2= -0.07$ are chosen. }}
 	\label{f5a}
 \end{figure}

\begin{figure}[H]
 	\centering
 	{\includegraphics[width=.45\textwidth]{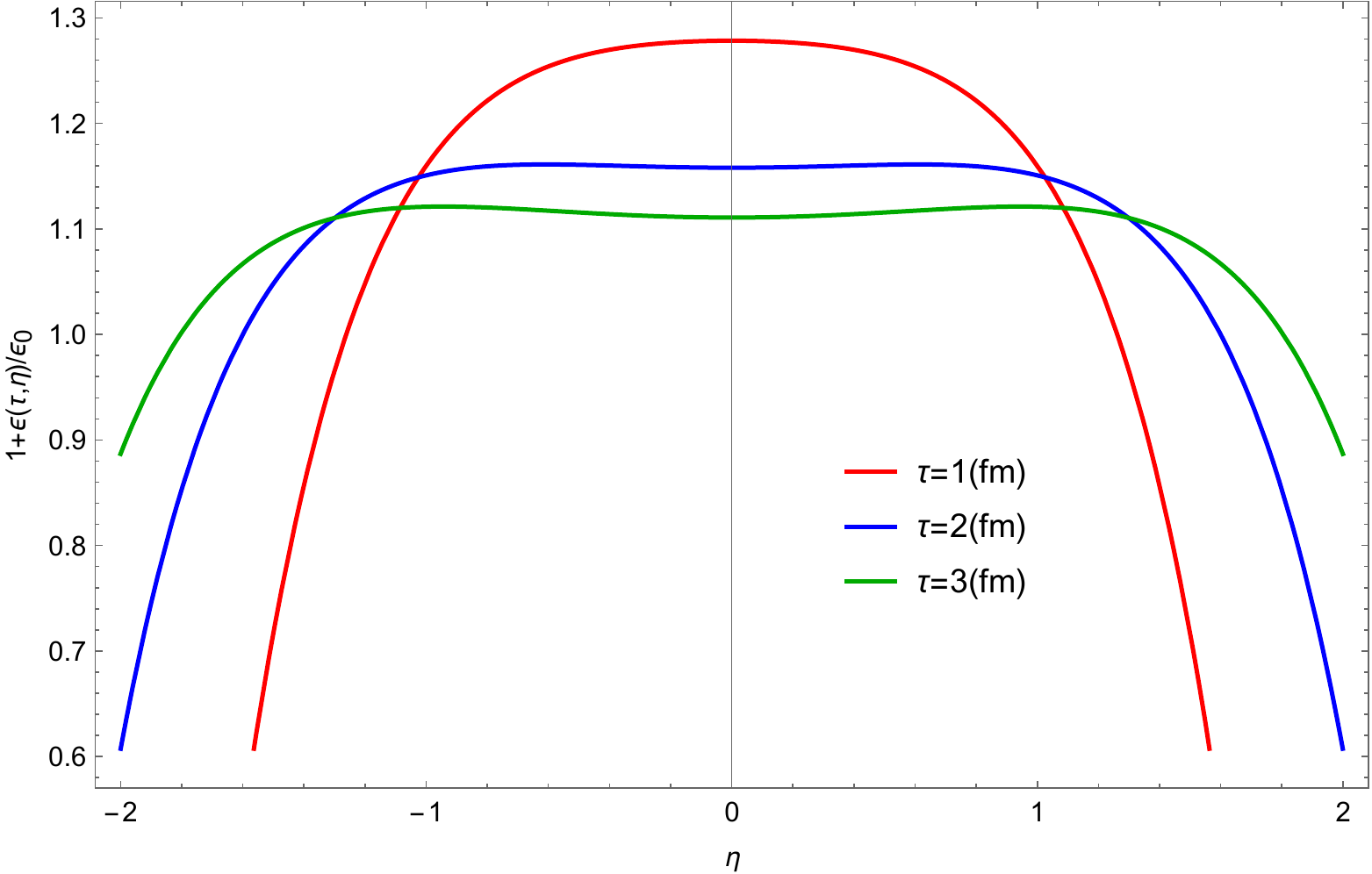}}
 	{\includegraphics[width=.45\textwidth]{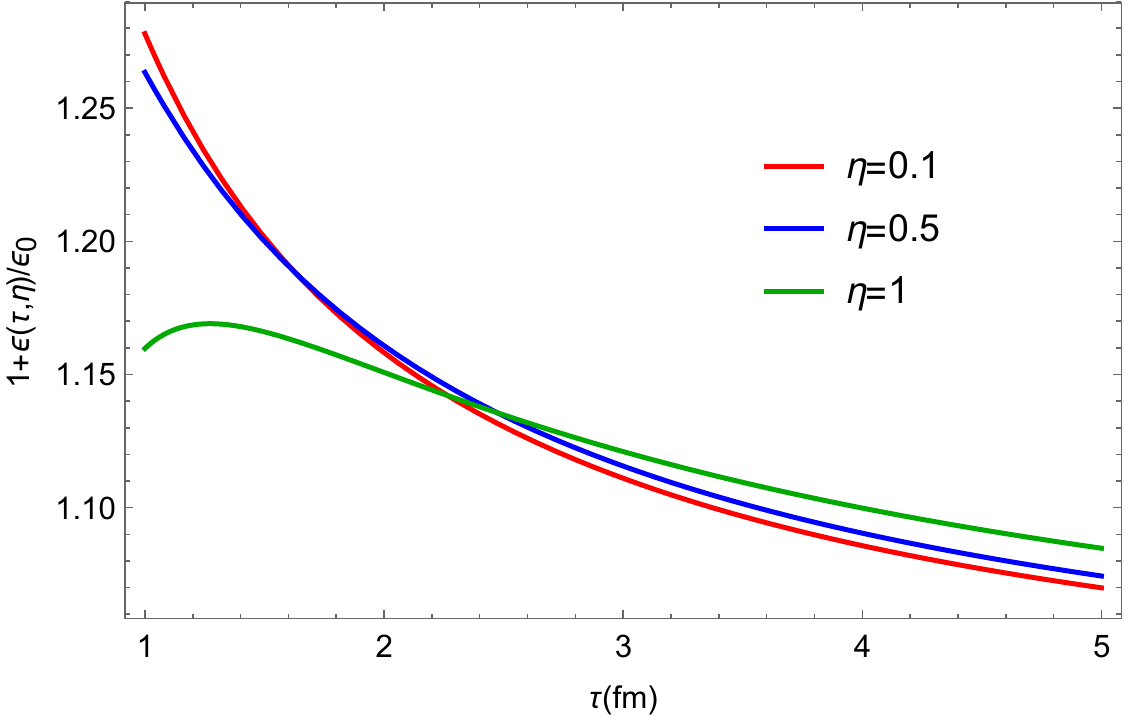}}
 	{\includegraphics[width=.45\textwidth]{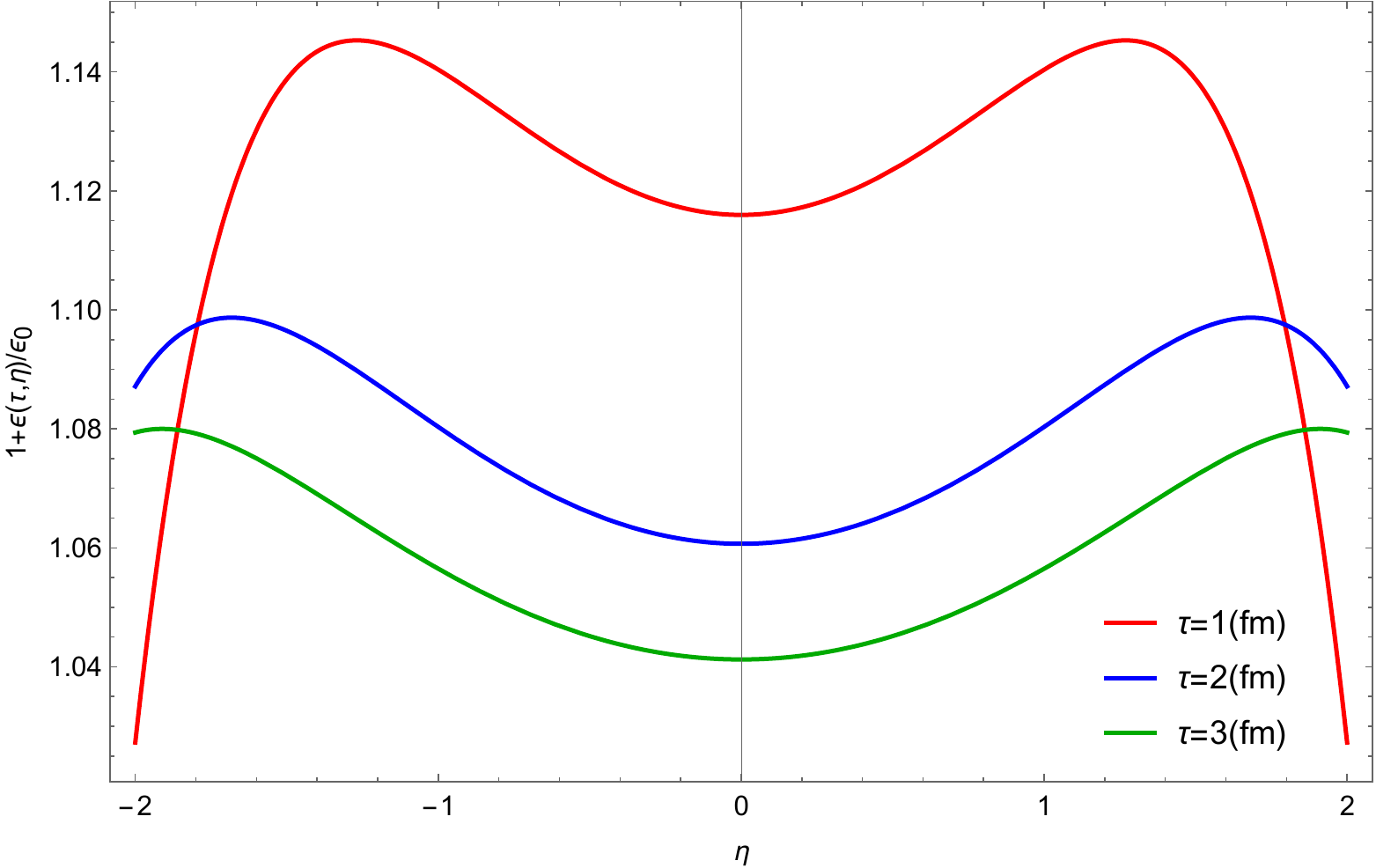}}
 	{\includegraphics[width=.45\textwidth]{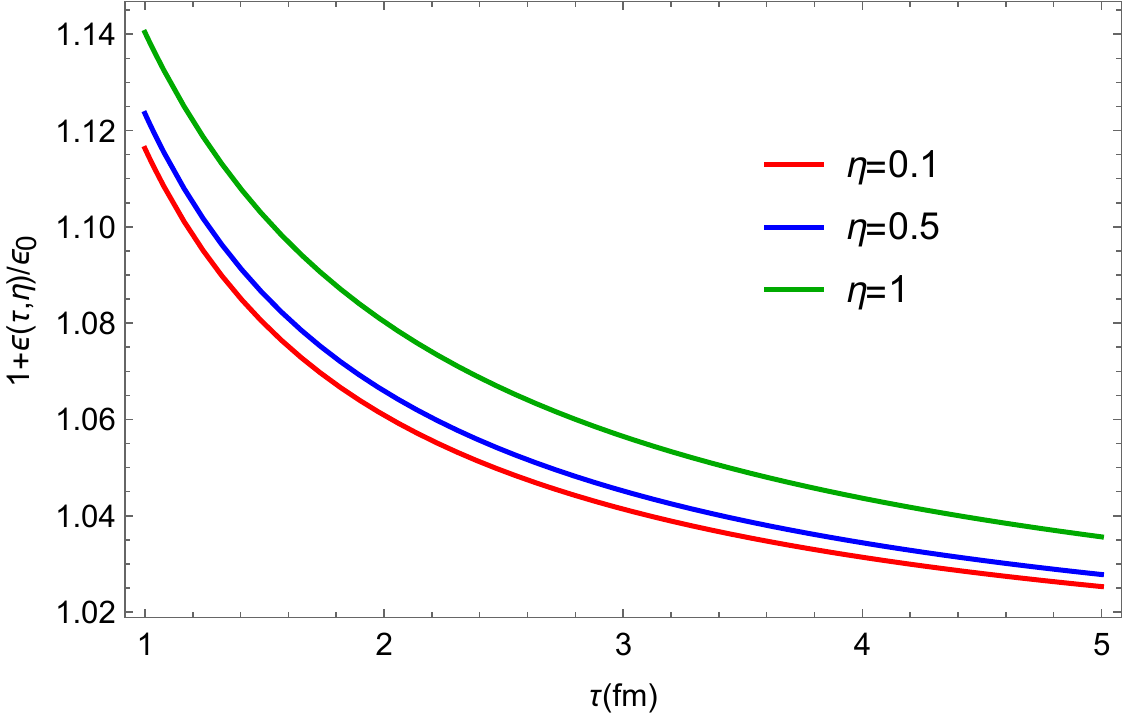}}
 	{\includegraphics[width=.45\textwidth]{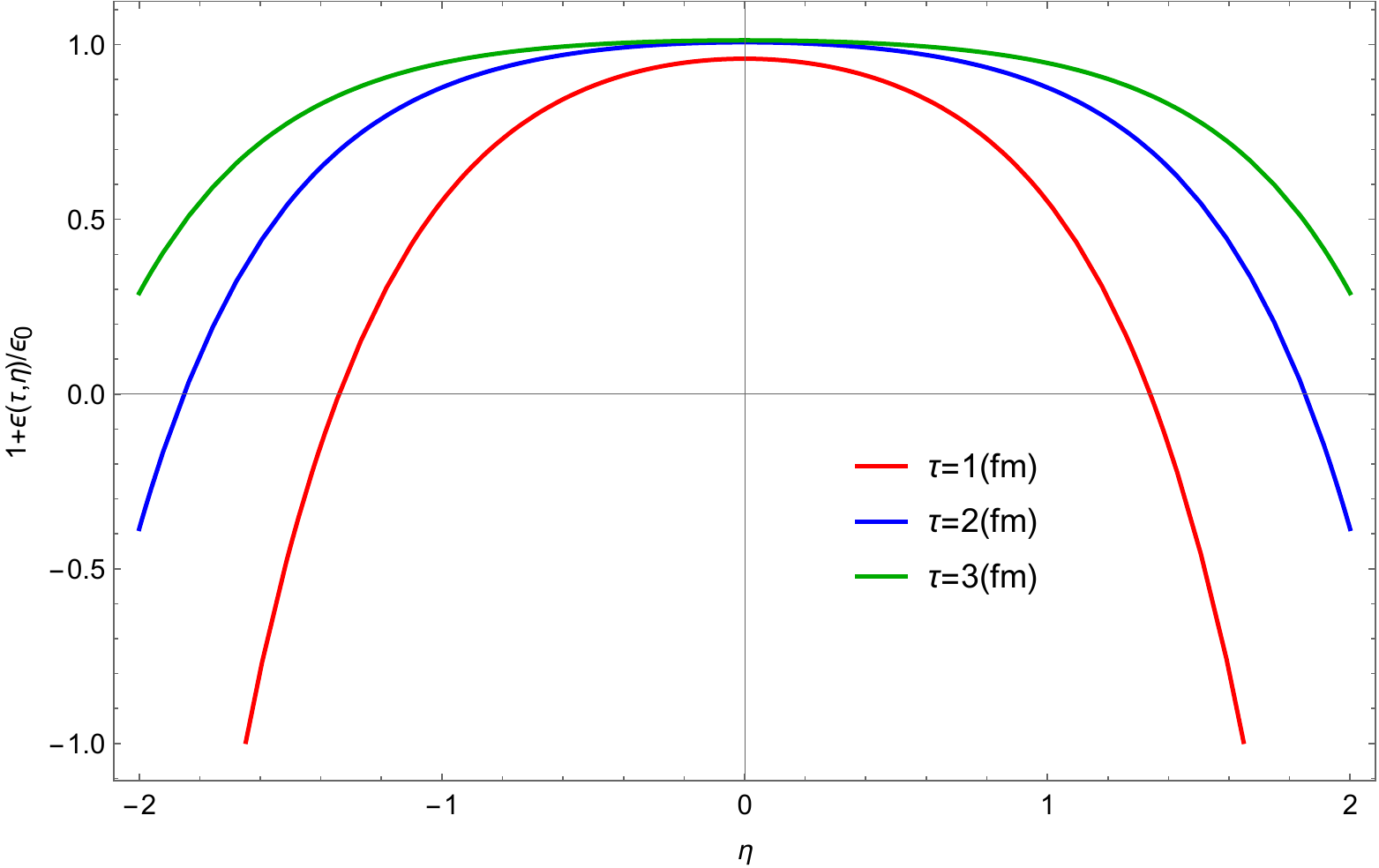}}
 	{\includegraphics[width=.45\textwidth]{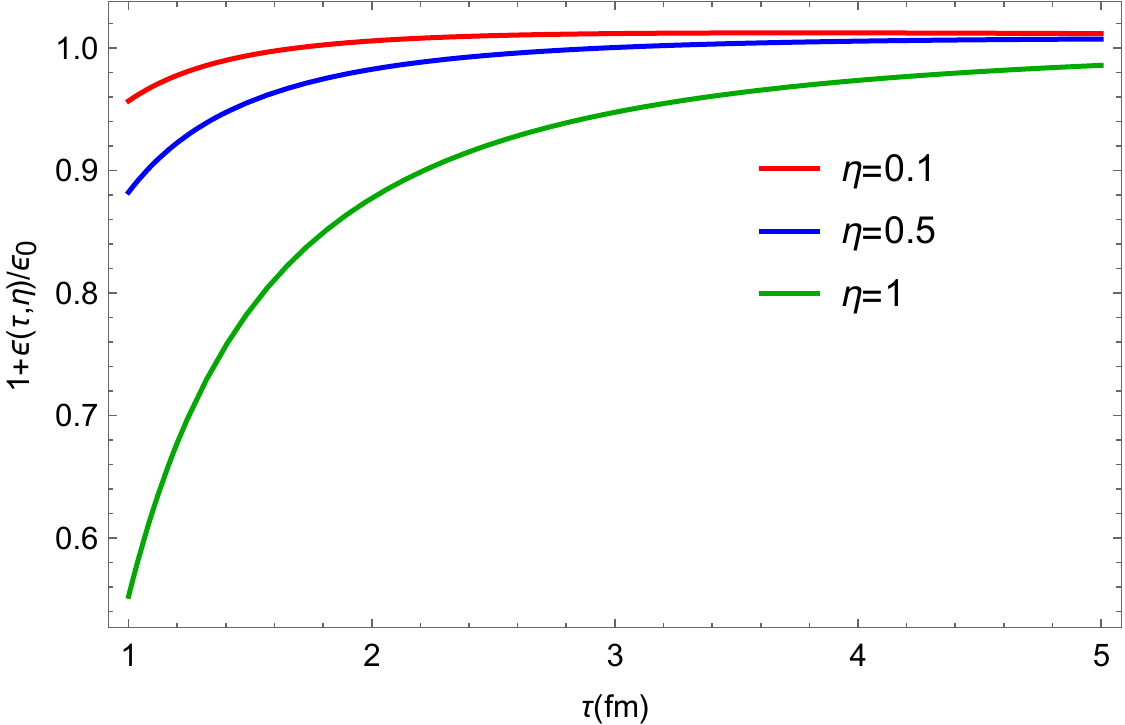}}		
 	\caption{\small{Left: $\big(\epsilon_\eta(\tau,\eta)+\epsilon_0\big)/\epsilon_0$ as a 	 function of   the space-time rapidity $\eta$ with different values of  $\tau$. Right: $\big(\epsilon_\eta(\tau,\eta)+\epsilon_0\big)/\epsilon_0$ as a function of proper time $\tau$ with different values of $\eta$.  The values $A_1, A_2$ are chosen: $ A_1 = 0.3, A_2= -0.07$(top panel),  $ A_1 = 0.1, A_2= -0.01$ (middle), and $ A_1 = 0.1, A_2= -0.1$ (bottom) respectively.			
 	}}
 	\label{f6}
 \end{figure}

Fig.~(\ref{f5a}) shows  the ratio of energy density $\epsilon(\tau,\eta)/\epsilon_0$  in terms of $\eta$ for
several values of $\tau$ ( left panel), or in terms of $\tau$ 
for several values of $\eta$ ( right panel). In according to the left panel of Fig.~(\ref{f5a}), the energy density rapidity distribution  at the early time when QGP is formed  has a Gaussian shape, while at the late time it becomes rather a plateau. It is found
that energy density slowly flows toward high rapidity at the
later time. 
The ratio of energy density $\epsilon(\tau,\eta)/\epsilon_0$  in terms of
 $\eta$   (the left panel) or in terms of   proper time $\tau$ ( right panel)  is exhibited in Fig.~(\ref{f6}) for
several different values of $A_1$ and $A_2$. As is evident, the shape of  energy density profile depends on the free parameters $A_1,A_2$. These parameters should be fixed with the rapidity dependence of particle yield, which can be computed in the theory and measured in experiments.
 Also, in left panel of Fig.~(\ref{f6}), we demonstrate
$\epsilon(\tau,\eta)/\epsilon_0$  in terms of rapidity for different fixed proper times. It can be seen that at the early times,
the plot has a Gaussian distribution, while at the late time, it becomes a plateau around the small rapidity. 

\begin{figure}[H]
 	\centering
 	{\includegraphics[width=.40\textwidth]{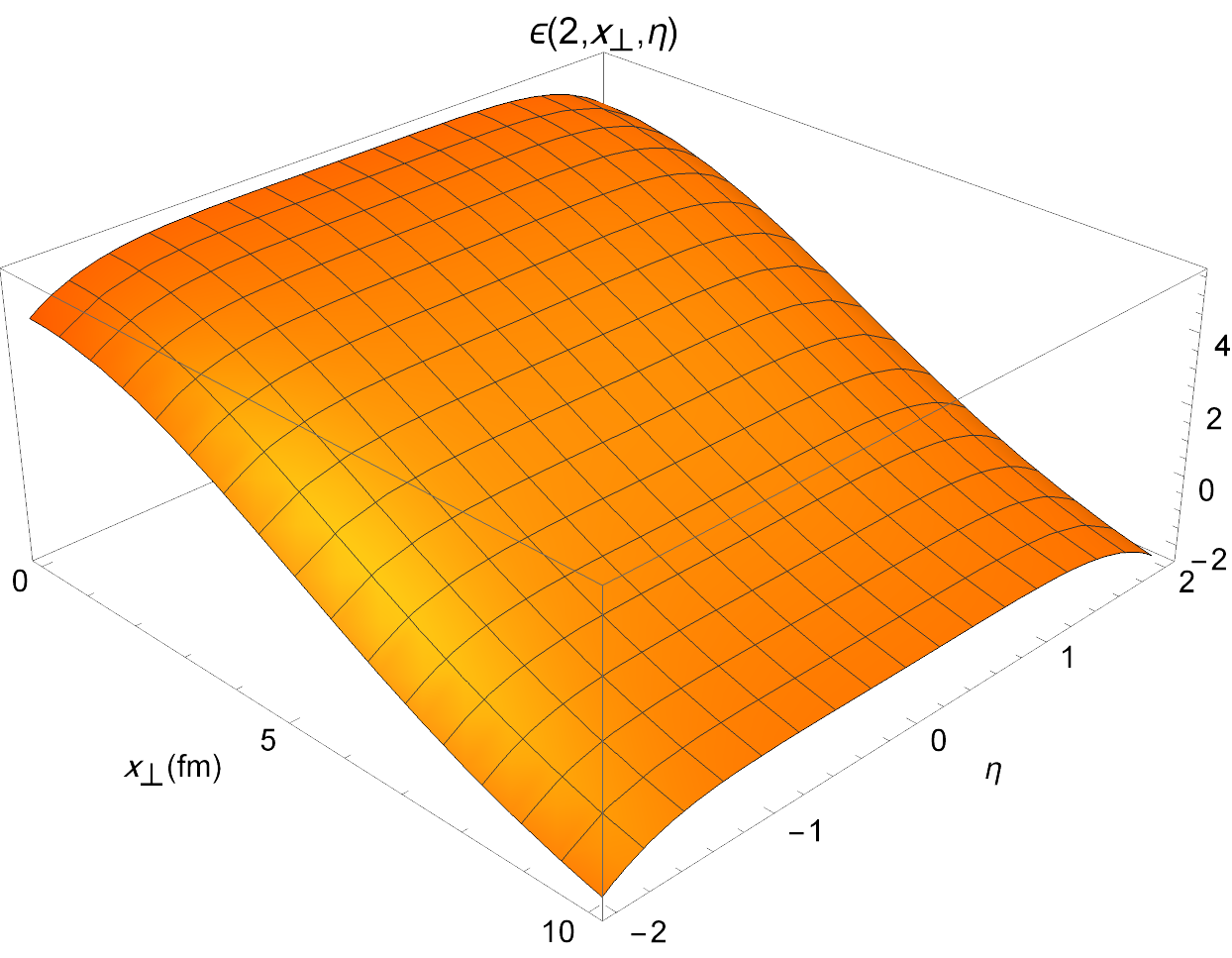}}
 	{\includegraphics[width=.40\textwidth]{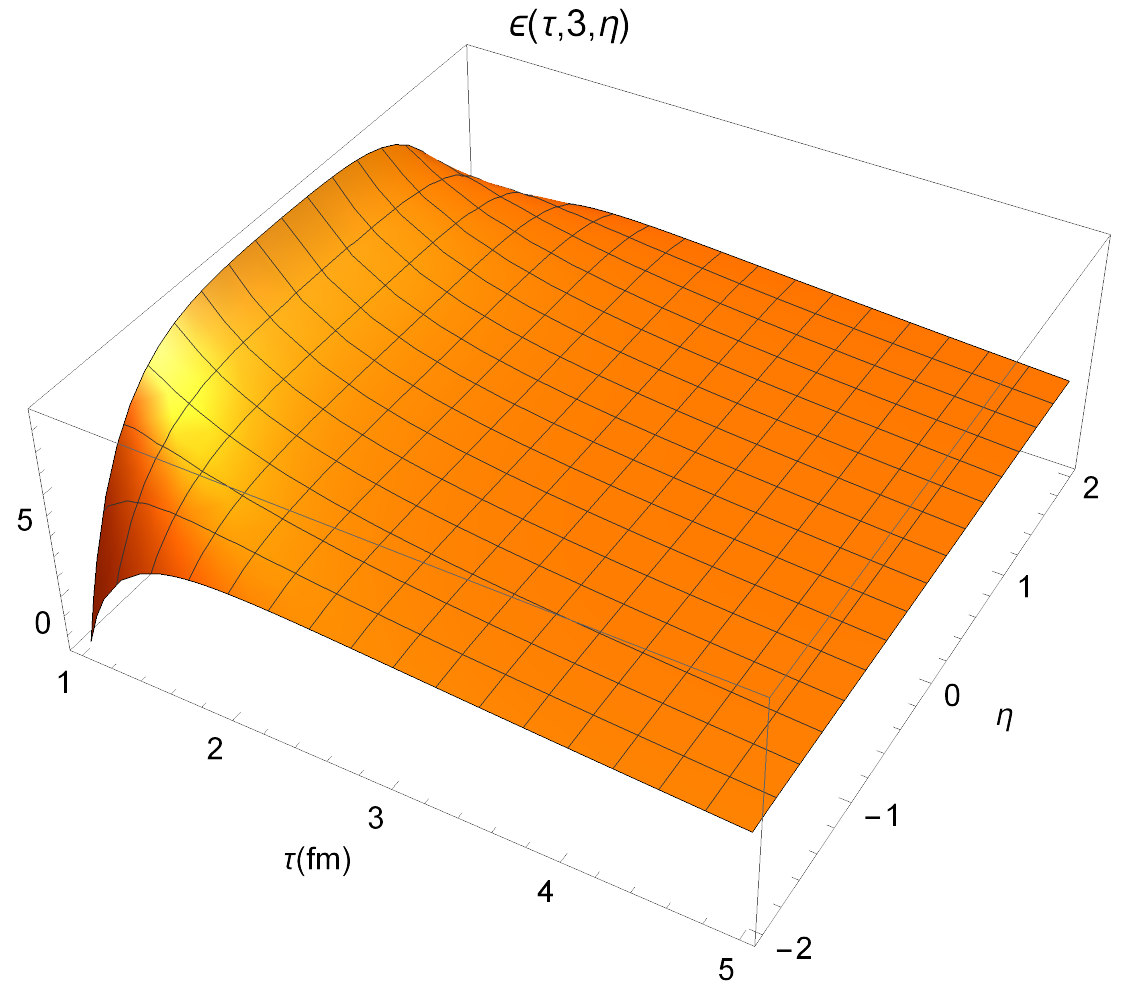}}
 	\caption{\small{$\epsilon(\tau,x_\perp,\eta)$.  The values $ A_1 = 0.3$ , $A_2= -0.07,$  $q=1/6.4(fm)^{-1}$ , $\tau_0=1 (fm)$, $\epsilon_c=5.4$ and $\hat{\epsilon}_0=1500$ are chosen. }}
 	\label{f7}
 \end{figure}

\begin{figure}[H]
 	\centering
 	{\includegraphics[width=.55\textwidth]{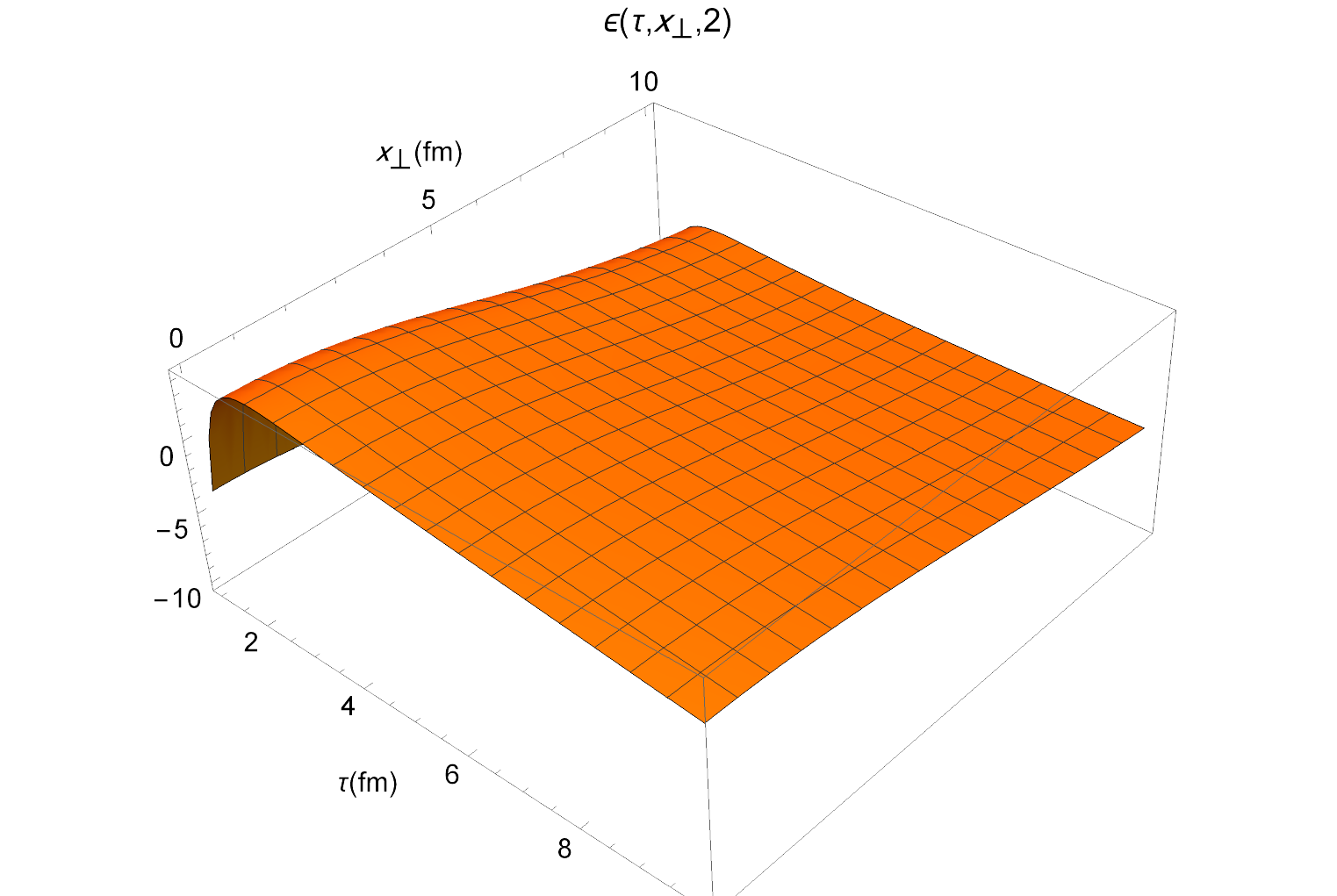}}
 	\caption{\small{$\epsilon(\tau,x_\perp,\eta)$.  The values $ A_1 = 0.3$ , $A_2= -0.07,$  $q=1/6.4(fm)^{-1}$ , $\tau_0=1 (fm)$, $\epsilon_c=5.4$ and $\hat{\epsilon}_0=1500$ are chosen. }}
 	\label{f8}
 \end{figure}

 Finally, Fig.~(\ref{f7}) illustrates the total energy density $\epsilon(\tau,x_\perp,\eta)$ as a function of $x_\perp$ and $\eta$ at a fixed value of $\tau=2 \ fm$, or as a function of $\tau$ and $\eta$ at a fixed value of $x_\perp=3 \ fm$. The figure indicates that the energy density is most significantly altered in the central region, where a reduction is observed in both cases. Additionally, Fig.~(\ref{f8}) displays the total energy density $\epsilon(\tau,x_\perp,\eta)$ as a function of $\tau$ and $x_\perp$ at a fixed value of $\eta=2$.
 

\section{Conclusion}
In this study, we present a novel extension of Bjorken flow that accounts for a medium with finite transverse dimensions, which undergoes radial and axial expansion. Additionally, we explicitly consider the breaking of boost invariance, while preserving its fundamental characteristics. The conservation equations are solved analytically and perturbatively, leading to the derivation of a new set of exact solutions for the 1 + 2D ideal hydrodynamics. These solutions are capable of accurately describing heavy-ion collisions at finite collision energies. We provide a detailed analysis of the modifications to the fluid velocity and energy density resulting from our proposed approach.

Our analysis is restricted to perfectly central collisions, necessitating the assumption of rotation invariance around the beam line. Consequently, all relevant quantities are solely dependent on the proper time $\tau$, transverse position $x_\perp$, and rapidity $\eta$ in the $ (\tau, x_\perp, \phi, \eta)$ coordinate system. Utilizing a first-order perturbation expansion, we derive the energy density, transverse flow velocity, and flow rapidity.

 In order to investigate the expansion of flow in the transverse plane, we have chosen to utilize the Gubser flow as our initial flow configuration. Subsequently, we proceed to compare the outcomes of our model with those of the Gubser flow.  Furthermore, we have assessed the transverse flow velocity obtained from our model in relation to the phenomenological proposition of $v_\perp=\frac{x_\perp}{50}$. Furthermore, an analysis was conducted to determine the distribution of energy density. The findings indicate that it is comparatively smoother than that of Qubser flow.
  
The current study aims to examine a model that demonstrates a partial breakdown of boost invariance in longitudinal expansion. To achieve this, power series expansions up to the first order in perturbation expansion were utilized. The investigation focused on the propagation of small perturbations and their impact on the correction of energy density and acceleration parameter. Analytical solutions were computed for the distribution of energy density in longitudinal expansion and the acceleration parameter of the fluid. By selecting appropriate values for the free parameters A1 and A2 in the solution, satisfactory results were obtained for the rapidity fluid velocity and energy density correction.

In our analysis, we have adopted a perturbative approach to model the plasma, which is superimposed on the background flow, rather than conducting a comprehensive hydrodynamical calculation. The insignificance of the observed effects validates our methodology. However, we must emphasize that our calculations are based on several crucial assumptions, namely: (i) we have considered the medium's two fundamental properties, assuming central collisions that lead to azimuth symmetry, a small transverse velocity compared to longitudinal expansion $(u_\perp << 1)$, and a soft breakdown of boost invariance $(Y-\eta<<1)$. Relaxing these assumptions could yield intriguing outcomes and warrant further investigation. Nonetheless, any deviation from these assumptions would render the calculation significantly more complex.


\begin{thebibliography}{99}

\bibitem{a1} R. Andrade, F. Grassi, Y. Hama, T. Kodama, O. Socolowski Jr.,``On the necessity to include event-by-event fluctuations in experimental evaluation of elliptical flow", Phys. Rev. Lett. 97, 202302 (2006).

\bibitem{a2} P. Romatschke, U. Romatschke,``Viscosity information from relativistic nuclear collisions: how perfect is the fluid observed at RHIC?", Phys. Rev. Lett. 99, 172301 (2007).

\bibitem{a3} H. Song, U.W. Heinz,``Causal viscous hydrodynamics in 2+1 dimensions for relativistic heavy-ion collisions", Phys. Rev. C 77, 064901 (2008).

\bibitem{a4} P. Bozek,``Flow and interferometry in 3+1 dimensional viscous hydrodynamics", Phys. Rev. C 85, 034901 (2012).

\bibitem{a5} C. Gale, S. Jeon, B. Schenke, P. Tribedy, R. Venugopalan,``Event-by-event anisotropic flow in heavy-ion collisions from combined Yang-Mills and viscous fluid dynamics", Phys. Rev. Lett. 110(1), 012302 (2013).

\bibitem{a6} L. Del Zanna, et al.,``Relativistic viscous hydrodynamics for heavy-ion collisions with ECHO-QGP", Eur. Phys. J. C 73, 2524 (2013).

\bibitem{a1new}    J.D. Bjorken, Highly relativistic nucleus–nucleus collisions: the central rapidity region. Phys. Rev. D 27, 140 (1983).


\bibitem{a7}  S.S. Gubser, Symmetry constraints on generalizations of Bjorken
flow. Phys. Rev. D 82, 085027 (2010).	

\bibitem{a8} T. Csorgo, M.I. Nagy, M. Csanad, A new family of simple solutions of relativistic perfect fluid Hydrodynamics.
Phys. Lett. B 663, 306 (2008).

\bibitem{a9}      K.J. Eskola, K. Kajantie, P.V. Ruuskanen, Hydrodynamics
of nuclear collisions with initial conditions from perturbative
QCD, Eur. Phys. J. C 1, 627-632 (1998).
 \bibitem{a10}      P. Bozek, I. Wyskiel, Rapid hydrodynamic expansion in relativistic
heavy-ion collisions, Phys. Rev. C 79, 044916 (2009).
 \bibitem{a11}  P. Bozek, Viscous evolution of the rapidity distribution of matter
created in relativistic heavy-ion collisions, Phys. Rev. C 77,
034911 (2008).

\bibitem{a12} J. Ze-Feng, Y. Chun-Bin, Mate Csanad and Tamas Csorgo,
” Accelerating hydrodynamic description of pseudorapidity
density and initial energy density in p + p, Cu+Cu, Au+Au
and Pb+Pb collisions at energies available at BNL Relativistic
Heavy Ion Collider and the CERN Large Hadron Collider,
Phys. Rev. C 97, 064906 (2018) ”.

\bibitem{aa12} Shuzhe Shi, Sangyong Jeon, and Charles Gale, ”Family of new exact solutions for longitudinally expanding ideal fluids” Phys. Rev. C 105, L021902 (2022).


\bibitem{a13}  M. I. Nagy, T. Csörgő, and M. Csanád, " Detailed description of accelerating, simple solutions of relativistic perfect fluid hydrodynamics", Phys. Rev. C 77, 024908        (2008).

\bibitem{a14}   T. Csorgo, M. I. Nagy, and M. Csanad, A New family
of simple solutions of perfect 
fluid hydrodynamics, Phys. Lett. B 663, 306 (2008).
\bibitem{a15}    S. Amai, H. Fukuda, C. Iso, and M. Sato, Hydrodynamical
treatment of multiple meson production in high energy
nucleon-nucleus collisions, Progress of Theoretical
Physics 17, 241 (1957).
\bibitem{a16} A. Bialas, R. A. Janik, and R. B. Peschanski, Unified
description of Bjorken and Landau 1+1 hydrodynamics,
Phys. Rev. C 76, 054901 (2007).
\bibitem{a17} G. Beuf, R. Peschanski, and E. N. Saridakis, Entropy 
flow
of a perfect 
fluid in (1+1) hydrodynamics, Phys. Rev. C 78, 064909 (2008).
\bibitem{a18}    T. Mizoguchi, H. Miyazawa, and M. Biyajima, A Potential including Heaviside function in 1+1 dimensional hydrodynamics by Landau, Eur. Phys. J. A 40, 99 (2009).
\bibitem{a19}       R. Peschanski and E. N. Saridakis, Exact (1+1)-dimensional flows of a perfect 
fluid, Nucl. Phys. A 849, 147 (2011).
\bibitem{a20} C.-Y. Wong, A. Sen, J. Gerhard, G. Torrieri, and
K. Read, Analytical Solutions of Landau (1+1)-
Dimensional Hydrodynamics, Phys. Rev. C 90, 064907
(2014).







 \bibitem{a22} M. Greif, C. Greiner, Z. Xu,``Magnetic field influence on the early time dynamics of heavy-ion collisions", Phys. Rev. C 96, 014903 (2017).

\bibitem{a23} M. H. Moghaddam, B. Azadegan, A. F. Kord, W. M. Alberico,``Non-relativistic approximate numerical
ideal-magneto-hydrodynamics of (1+1D) transverse flow in Bjorken scenario", Eur. Phys. J. C 78, 255 (2018).

\bibitem{a24} G. Inghirami, M. Mace, Y. Hirono, L. Del Zanna, D. E. Kharzeev, M. Bleicher,``Magnetic fields in heavy-ion collisions: flow and charge transport", Eur. Phys. J. C 80, 293 (2020).
\bibitem {aa25}  Ankit Kumar Panda, Ashutosh Dash, Rajesh Biswas and Victor Roy, "Relativistic non resistive 
 viscous magnetohydrodynamics from the kinetic theory: a relaxation time approach", JHEP 03, 216 (2021).
\bibitem{aa26} Ankit Kumar Panda, Ashutosh Dash, Rajesh Biswas and Victor Roy, Relativistic resistive
dissipative magnetohydrodynamics from the relaxation time approximation, Phys. Rev. D
104, 054004 (2021).
\bibitem{aa27} GS Denicol, E Moln´ar, H Niemi, DH Rischke , Resistive dissipative magnetohydrodynamics
from the Boltzmann-Vlasov equation, Phys. Rev. D 99, 056017 (2019).

\bibitem{a25}    P. F. Kolb and R. Rapp, Transverse 
fow and hadro-chemistry in Au + Au collisions at $\sqrt{s_{NN}}= 200 \  GeV$," Phys. Rev. C67  044903 (2003).

\bibitem{a33}  PHENIX Collaboration, K. Adcox et. al., "Formation of dense partonic matter in relativistic nucleus nucleus collisions at rhic: Experimental evaluation by the phenix
collaboration," Nucl. Phys. A757  184-283  (2005). 
\bibitem{a34} D. Teaney,``Effect of shear viscosity on spectra, elliptic flow, and Hanbury Brown–Twiss radii", Phys. Rev. C 68, 034913 (2003).
\bibitem{a35} Gubser SS. Symmetry constraints on generalizations of Bjorken flow. Physical Review D. 2010 Oct 26;82(8):085027.

\bibitem{a26}   U. Gursoy, D. Kharzeev, K. Rajagopal, Magnetohydrodynamics,
charged currents and directed flow in heavy ion collisions. Phys.
Rev. C 89(5), 054905 (2014).










\end{thebibliography}
\end{document}